\documentclass[journal,draftcls,onecolumn,12pt,twoside]{IEEEtran}

\usepackage{amssymb,color}
\usepackage{amsmath}
\usepackage{bbm}
\usepackage{graphicx}
\usepackage{epstopdf}
\usepackage{amsthm,amsfonts}
\usepackage{tikz}

\usepackage{subcaption}

\usepackage{accents}

\usepackage{algorithm}
\usepackage{algorithmic}
\usepackage{psfrag}
\usepackage{cite}
\usepackage{pifont}
\usepackage[font={small}]{caption}

\allowdisplaybreaks

\makeatletter
\newcommand{\oset}[3][0ex]{%
  \mathrel{\mathop{#3}\limits^{
    \vbox to#1{\kern-2\ex@
    \hbox{$\scriptstyle#2$}\vss}}}}
\makeatother

\begin{document}
\title{Integrated Sensing and Communication for Large Networks using Joint Detection and a Dynamic Transmission Strategy} 

\author{
\IEEEauthorblockN{\large  Konpal Shaukat Ali and Marwa Chafii}

\thanks{The authors are with the Engineering Division, New York University (NYU) Abu Dhabi, 129188, UAE (Email: \{konpal.ali, marwa.chafii\}@nyu.edu). M. Chafii is also with NYU WIRELESS, NYU Tandon School of Engineering, Brooklyn, 11201, NY.

 }}

\maketitle

\begin{abstract} 
A large network employing integrated sensing and communication (ISAC) where a single transmit signal by the base station (BS) serves both the radar and communication modes is studied. We consider bistatic detection at a passive radar and monostatic detection at the transmitting BS. The radar-mode performance is significantly more vulnerable than the communication-mode due to the double path-loss in the signal component while interferers have direct links. To combat this, we propose: 1) a novel dynamic transmission strategy (DTS), 2) joint monostatic and bistation detection via cooperation at the BS. We analyze the performance of monostatic, bistatic and joint detection. {We show that bistatic detection with dense deployment of low-cost passive radars offers robustness in detection for farther off targets.} Significant improvements in radar-performance can be attained with joint detection in certain scenarios, while using one strategy is beneficial in others. Our results highlight that with DTS we are able to significantly improve quality of radar detection at the cost of quantity. Further, DTS causes some performance deterioration to the communication-mode; however, the gains attained for the radar-mode are much higher. We show that joint detection and DTS together can significantly improve radar performance from a traditional radar-network.

\end{abstract}

\IEEEpeerreviewmaketitle

\section{Introduction}

There is an ever increasing need for connectivity which in next generation networks will not only mean growing communication requirements like higher throughput and reliability but also increased sensing requirements. A number of applications such as vehicle-to-vehicle communication, drones, smart cities and more generally the internet of things (IoT) have both high communication and sensing requirements \cite{jrc_overview0,jrc_overview1}. Although radar sensing and communication share a lot in terms of both software and hardware, for decades, the two have been advancing independently with limited intersection \cite{jrc_overview0,jrc_overview1}. Since the wireless spectrum is scarce, it is natural to consider systems that are able to `reuse' the spectrum more efficiently and utilize it for both communication and sensing simultaneously.

{Two approaches have been pursued in the context of having radar and communication share spectrum, namely \emph{coexistence} and \emph{co-design} \cite{jrc_sg6,12challenges}. Under coexistence, different transmit signals and therefore hardware is used for the communication and radar modes. Under co-design, a \emph{single} signal is used for both modes. In general, these systems have been referred to by many names such as joint radar and communication (JRC), joint communication and radar (JCR), joint communication and radar/radio sensing (JCAS), radar-communication (RadCom), dual-function radar communications (DFRC) and \emph{integrated sensing and communication} (ISAC) \cite{jrc_overview0}. ISAC and DFRC have frequently been used to refer to a system that jointly designs and uses a \emph{single} transmit signal for both communication and sensing, i.e., co-design; in this work we use the term ISAC to study such a system.} ISAC offers improved spectral efficiency and reduced hardware costs \cite{jrc_sg4}. Further, in contrast to coexisting communication and sensing systems where a single transmit signal does not serve both the radar and communication-modes, ISAC also protects the network from additional interference.


The majority of work on ISAC has focused on setups with a single-cell/few cells \cite{jrc_irs1,Bazzi_JRC1,Bazzi_JRC2}. Since real networks are becoming very dense and have high interference, studying setups with a single-cell/few cells can lead to inaccurate deductions as well as result in suboptimum parameter selection that can hurt performance in a real network \cite{my_nomaMag}. Studying a large network that mimics a real-world network is thus of great value. Stochastic geometry provides a unified mathematical paradigm for modeling large wireless networks and characterizing their operation while taking intercell interference into account \cite{MH_Book2,h_tut,myNOMA_tcom,mySecrecy,myPartialNOMA}.

 Works such as \cite{jrc_sg1,jrc_sg2,jrc_sg3,jrc_sg4} have studied coexisting communication and sensing using stochastic geometry tools; however, these setups {do not focus on co-design of the two modes} {as the same transmit signal is not used by {both} modes simultaneously}. In \cite{jrc_sg1}, a network where nodes share a channel to work in the radar and communication-modes in different time slots is studied. The impact of only the strongest interferer is considered, thus the interference from the entire network is not accounted for. In \cite{jrc_sg2}, multi-radar cooperative detection is used to enhance the detection range by clusters of radars sharing their sensing information via the communication-mode. It is also assumed that the radar-mode does not receive interference from both the communication and radar-modes. In \cite{jrc_sg3}, a 1D setup of vehicles is studied where communication and radar are on different bandwidths and the goal is to use cooperation between vehicles to increase the detection range. Interference from the network has not been considered. In \cite{jrc_sg4}, a bistatic setup is studied where the transmitting base station (BS) and detecting radar are at fixed locations while multiple users (UEs) and clutter scatterers are distributed according to a Poisson point process (PPP). The BS serves the radar and communication functionalities in a time division manner. 
 
{On the other hand, the works in \cite{jrc_sg5,jrc_sg6} study systems where there is co-design of the radar and communication modes in a large network. While \cite{jrc_sg6} is said to focus on co-design, a setup where the vehicle of interest receives a communication signal from one vehicle and a radar signal from a different vehicle is studied; further, the receiver for each mode on the vehicle of interest is separate. The impact of interference cancellation on the two modes is analyzed and it is found that interference cancellation at the radar mode is much more beneficial than that at the communication mode. Note that while stochastic geometry tools are used to model the 1D locations of vehicles on two lanes, interference from only the nearest interferer is considered; {thus the impact of interference from a large network is not accounted for}. In \cite{jrc_sg5}, BSs use the same signal to send communication messages and detect targets; however, the double path loss associated with target detection has not been taken into account. The potential spectral efficiency of ISAC networks is studied and used to calculate the energy efficiency. It is found that a BS density that optimizes energy efficiency exists and is different from the optimum BS density in the case of a traditional communication network. Different from the works in \cite{jrc_sg5,jrc_sg6}, we study the performance in a large network where the impact of interference from the entire network and the double path loss associated with target detection in the radar mode are both accounted for. }

The radar-mode is particularly vulnerable as its signal component \textcolor{red}{suffers a double path loss} due to the nature of radar detection. Further, in a large network, the receiver in the radar-mode still incurs interference from all the nodes in the network directly, which do not undergo a double path loss effect. These two factors play a significant role in deteriorating the signal to interference and noise ratio (SINR) of the radar-mode. The communication-mode, on the other hand, is `safe' as it does not incur deterioration compared to a regular cellular network without ISAC.



{Radar sensing can be done via monostatic detection, i.e., the transmitter of the signal detects the echo, if any, as well as via bistatic detection where the transmitter is not collocated with the radar receiver. In the ISAC network, the cellular BS transmits a signal that is also used by the radar-mode. Bistatic detection seems promising in such networks since it is convenient to deploy many low-cost passive radars for listening, as the radar-mode utilizes the communication-mode's spectrum as well as the transmit power. Monostatic detection at the BS is also used in many works such as \cite{jrc_sg5,Bazzi_JRC1,Bazzi_JRC2}; this, however, requires full duplex (FD) operation as the BS simultaneously transmits and receives on the same channel. While simultaneously transmitting and receiving messages on the same channel was not possible previously due to the overwhelming self-interference (SI), with advances in transceiver design, successful SI cancellation (SIC) that enables FD operation is now possible. Thus, using SIC, a transceiver is able to simultaneously transmit and receive on the same frequency channel, i.e., operate in FD \cite{2sic2_fd5,1sic1_fd5}. A number of works exist on FD \cite{myD2DFD} and in \cite{FD_JRCmag} FD operation is identified as a key enabler for such ISAC systems.} 



In this  work we study a large network where the radar and communication modes share a single transmit signal and operate in the same frequency channel simultaneously. {We take into account path loss, fading and intercell interference from the large network. We study the impact of both FD monostatic and bistatic detection. In light of the significant deterioration to the SINR of the radar-mode, we propose two solutions to improve the radar-mode's performance without significantly hurting the communication-mode: 1) a novel \emph{dynamic transmission strategy (DTS)} for ISAC, 2) joint bistatic and FD monostatic detection. {Since the cellular BS has strong compute power and back-haul connections, the use of joint detection which involves cooperation between the bistatic and monostatic detection is possible; thus, detection by either one of the approaches results in successful detection of the radar-mode.} To the best of our knowledge, this is the first work to study ISAC in a large network taking into account the impact of interference and the double path loss associated with target detection. Further, this is the first work to propose and study joint bistatic and monostatic detection {as well as the first to propose a strategy like the DTS for ISAC}.} The contributions of this work can be summarized as follows:
\begin{itemize}
\item {We study the impact on the radar-mode performance of bistatic and monostatic detection and propose using joint detection to enhance performance in a large network where ISAC technology is deployed. } 
\item The DTS is proposed as a novel strategy where each BS dynamically transmits with high power in one of $M$ slots while the remaining slots have lower power transmissions. While communication takes place in all $M$ slots, by limiting radar-mode detection to the slot with high power, both the signal is improved and relative interference power is reduced for the radar-mode. This way we trade the \emph{quantity} of radar-mode detection for \emph{quality}. 
\item We provide a tractable analytical framework to quantify the performance of the communication-mode, the radar-mode with bistatic detection, with monostatic detection, as well as with joint detection. 
\item We identify scenarios, such as cell-center targets, where monostatic detection is superior to bistatic detection, and other scenarios such as those of farther off targets where bistatic detection is superior. Further, we show that while joint detection always improves performance, when one of bistatic and monostatic detection is not significantly superior to the other, using the joint detection results in non-trivial performance improvement. 
\item {We show that joint detection can improve the robustness to variation in performance with link distances.}
\item We find that while, as anticipated, the DTS hurts the communication-mode due to increased interference in the low power slots, this deterioration is much less than the gains attained by the radar-mode from DTS.
\item We show that with careful DTS parameter selection, the deterioration of the communication-mode can be reduced while improving the performance of the radar-mode. {In fact, interestingly, in certain scenarios, two choices of DTS parameters result in the same radar-mode performance while one of them results in superior communication-mode performance again highlighting the significance of parameter selection.}
\item We benchmark the performance of the ISAC network's radar-mode with a traditional radar-only network that does not use ISAC. We find that the DTS and joint detection significantly improve performance from that of a traditional radar-only network. These are thus important solutions to improve the performance of the radar-mode in ISAC. 
\end{itemize}


The rest of the paper is organized as follows: Section II describes the system model. The proposed DTS and methodology of analysis is in Section III. The SINR analysis is in Section IV. Section V presents the results and Section VI
concludes the paper.

\textit{Notation:} We denote vectors using bold text, $\|\textbf{z}\|$ is used to denote the Euclidean norm of the vector $\textbf{z}$. The ordinary hypergeometric function is denoted by ${}_2F_1$. The cdf (ccdf, pdf) of the RV $X$ is denoted by $F_X$ ($\bar{F}_X$, $f_X$). The Laplace transform (LT) of the pdf of the RV $X$ is denoted by $\mathcal{L}_X(s)=\mathbb{E}[e^{-sX}]$.

\section{System Model}\label{SysMod}

\begin{figure}[htb]
\begin{minipage}[htb]{0.47\linewidth}
\centering\includegraphics[width=0.9\linewidth]{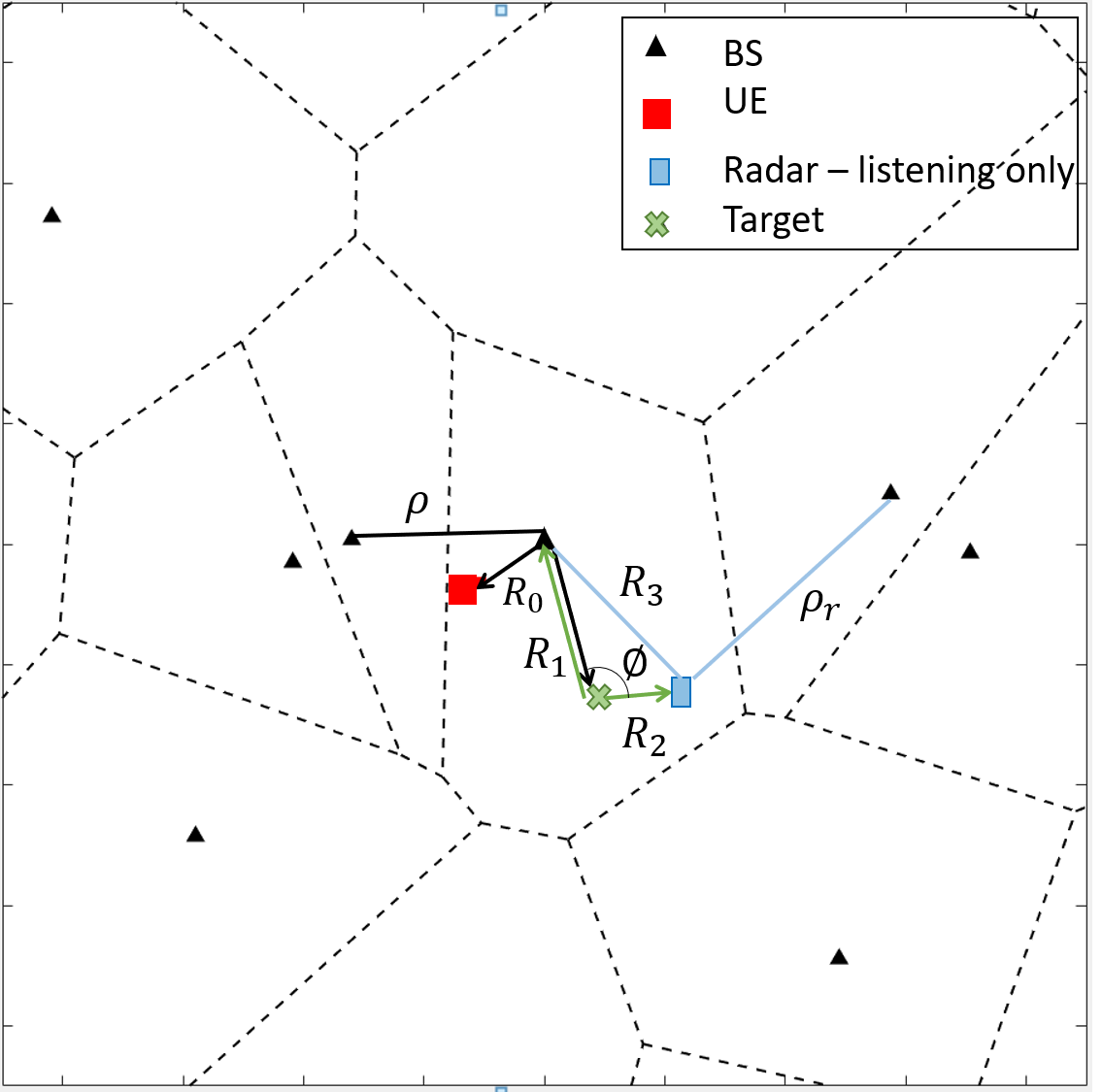}
\subcaption{Bistatic and {FD monostatic} ISAC}\label{framework1}
\end{minipage}\;\;\;
\begin{minipage}[htb]{0.47\linewidth}
\centering\includegraphics[width=0.9\linewidth]{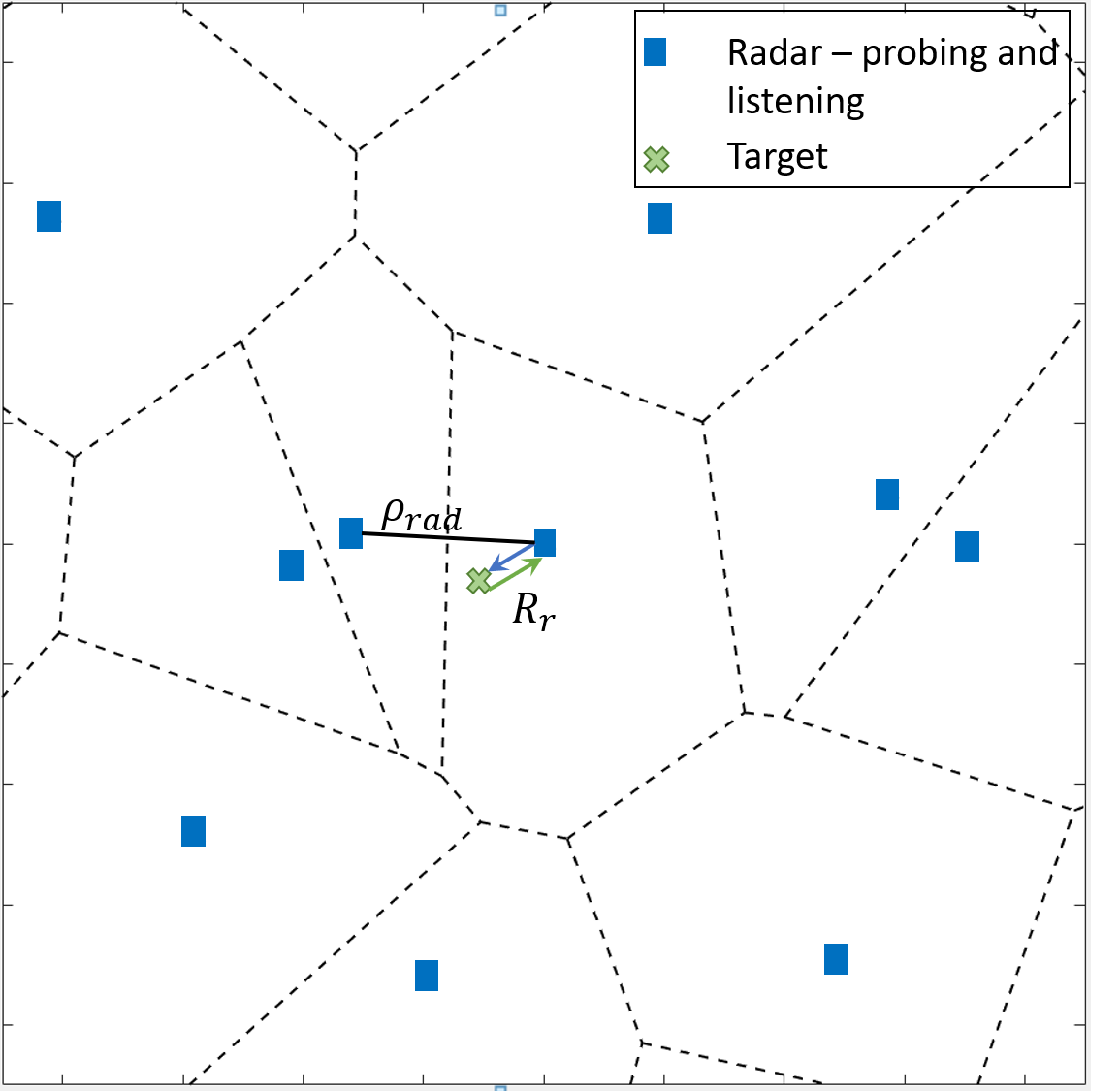}
\subcaption{Monostatic radar-only network}\label{frameworkR}
\end{minipage}
\caption{A snapshot realization of the network. All the nodes of interest in the typical cell have been shown.}\label{frameworksAll}
\end{figure}
\subsection{ISAC Network Model}
We consider a downlink cellular network where BSs are distributed according to a homogeneous PPP $\Phi$ with intensity $\lambda$. We assume communication-mode UEs are distributed uniformly and independently at random in the cells. {Additionally, there are radar-mode targets and radars present in the network. The radars can listen for the echo of the relevant target in the cell they are in}. In this work we consider both bistatic detection at the radar and monostatic detection at the BS via FD. The performance of bistatic and monostatic detection is analyzed separately and we also analyze the impact of joint detection via both techniques. A BS assigns each UE in its cell a unique time-frequency resource block to avoid interference between the messages of UEs in a cell. At the same time, in an effort to reuse the spectrum for sensing as well, the BS {assigns a target and a radar inside the cell to the resource block of a UE and} uses the communication signal to probe and sense the target at an unknown location inside the cell. This way, using ISAC, additional resources are not spent for detection in the radar-mode. Note that radars in the radar-mode do not transmit probing signals in the ISAC setup to avoid additional interference to the communication-mode; {thus} the radar-mode is allowed to use the communication-mode's spectrum for free without hurting the communication-mode. {We assume that the radar in bistatic detection has knowledge of the probing signal from the BS}; thus, a radar only detects targets inside the cell of the BS it is in. We also assume the BS is equipped with FD capability for monostatic detection.

To the network we add a BS at the origin $\textbf{o}$, which, under expectation over $\Phi$, becomes the typical BS (tBS) serving the typical UE (tUE) in the typical cell {via the communication-mode}. The echo of the tBS's message from the typical target (tTar) is listened for at the typical radar (tRad) in bistatic detection, at the tBS itself via FD in the case of monostatic detection {and at both in the case of joint detection}. In the remainder of this work, we study the performance of the typical cell. Since $\Phi$ does not include the BS at $\textbf{o}$, the set of interfering BSs for the tUE is $\Phi$. {As the tRad is aware of the message of the tBS, this is canceled and removed; thus, the set of interfering BSs for the tRad in the case of bistatic detection is also $\Phi$. Also, naturally, the set of interfering BSs at the tBS in the case of monostatic detection via FD is also $\Phi$.} {Since the focus of our work is not on FD operation, we assume perfect SIC at the tBS.} Fig. \ref{framework1} is a snapshot of the ISAC network where the nodes in the typical cell are shown. The tUE (tRad) lies a distance $R_0$ ($R_3$) from the tBS at $\textbf{o}$. The tTar lies a distance $R_1$ from the tBS at $\textbf{o}$. The tRad attempting to detect tTar lies a distance $R_2$ from it. {The distance between the tBS (tRad) and its nearest interferer is denoted by $\rho$ ($\rho_r$). }


\subsection{Radar-only Network Model}
In order to compare and benchmark the performance of our ISAC system, we introduce a radar-only network where the radars do not share communication spectrum. Here the radars are distributed according to a homogeneous PPP $\Phi_r$. {For fairness of comparison, $\Phi_r$ has the same intensity $\lambda$ as the PPP of BSs.} Monostatic radars are considered as it is assumed that radars cannot communicate with each other to cooperate due to a lack of the kind of infrastructure BSs have. Thus, a radar attempts to detect a target inside its cell by sending a probing signal with power $P_r$ and listening for its {reflected signal}. In contrast to the ISAC network, the radar-mode does not deal with interference from the PP of BSs; however, interference from the PP of radars sending probing signals exists. 

Similar to the ISAC setup, we add a radar at the origin \textbf{o}, which under expectation over $\Phi_r$ becomes the tRad searching for the tTar in the typical cell of the radar-only network. Since $\Phi_r$ does not include the radar at \textbf{o}, the set of interfering radars for the tRad is $\Phi_r$. Fig. \ref{frameworkR} shows a snapshot of the radar-only network. The tTar lies a distance $R_r$ from the tRad at \textbf{o}, and the distance between the tRad and its nearest interferer is denoted by $\rho_{\rm rad}$. 

\subsection{Channel Model and Link Distance Distribution}
We assume a Rayleigh fading environment such that the fading coefficient between any two nodes is i.i.d. with a unit mean exponential distribution. A power-law path loss model is considered where the signal decays at the rate $r^{-\eta}$ with distance $r$, $\eta>2$ denotes the path loss exponent and $\delta=\frac{2}{\eta}$.  


\subsubsection{ISAC Network} In the ISAC network the UEs are distributed uniformly and randomly in the cells and the BSs are distributed according to a PPP with intensity $\lambda$. Accordingly, the distribution of the link distance $R_0$ follows
\begin{align*}
f_{R_0} (x)=2\pi b \lambda x \exp(-\pi b \lambda x^2), \;\;\;\; x \geq 0,
\end{align*}
where $b=13/10$ is the correction factor due to the fact that the nodes of interest are in the typical cell, not in the 0-cell \cite[(12)]{mh_ue_PP}. 

We assume the tTar (tRad) lies a fixed distance $R_1=r_1$ ($R_2=r_2$) away from the tBS (tTar) in a random direction. The angle between the tBS-tTar link and the tTar-tRad link, as shown in Fig. \ref{framework1}, is denoted by $\phi$ and follows the distribution $f_{\phi}(u) =1/(2\pi)$, $0\leq u \leq 2\pi$. As a result, the distance between the tRad and tBS at $\textbf{o}$, denoted by $R_3$, is
\begin{align}
R_3=\sqrt{r_1^2 + r_2^2 - 2 r_2 r_1 \cos \phi}. \label{R3}
\end{align}

\begin{figure}[thb]
\begin{minipage}[htb]{0.47\linewidth}
\centering\includegraphics[width=0.9\linewidth]{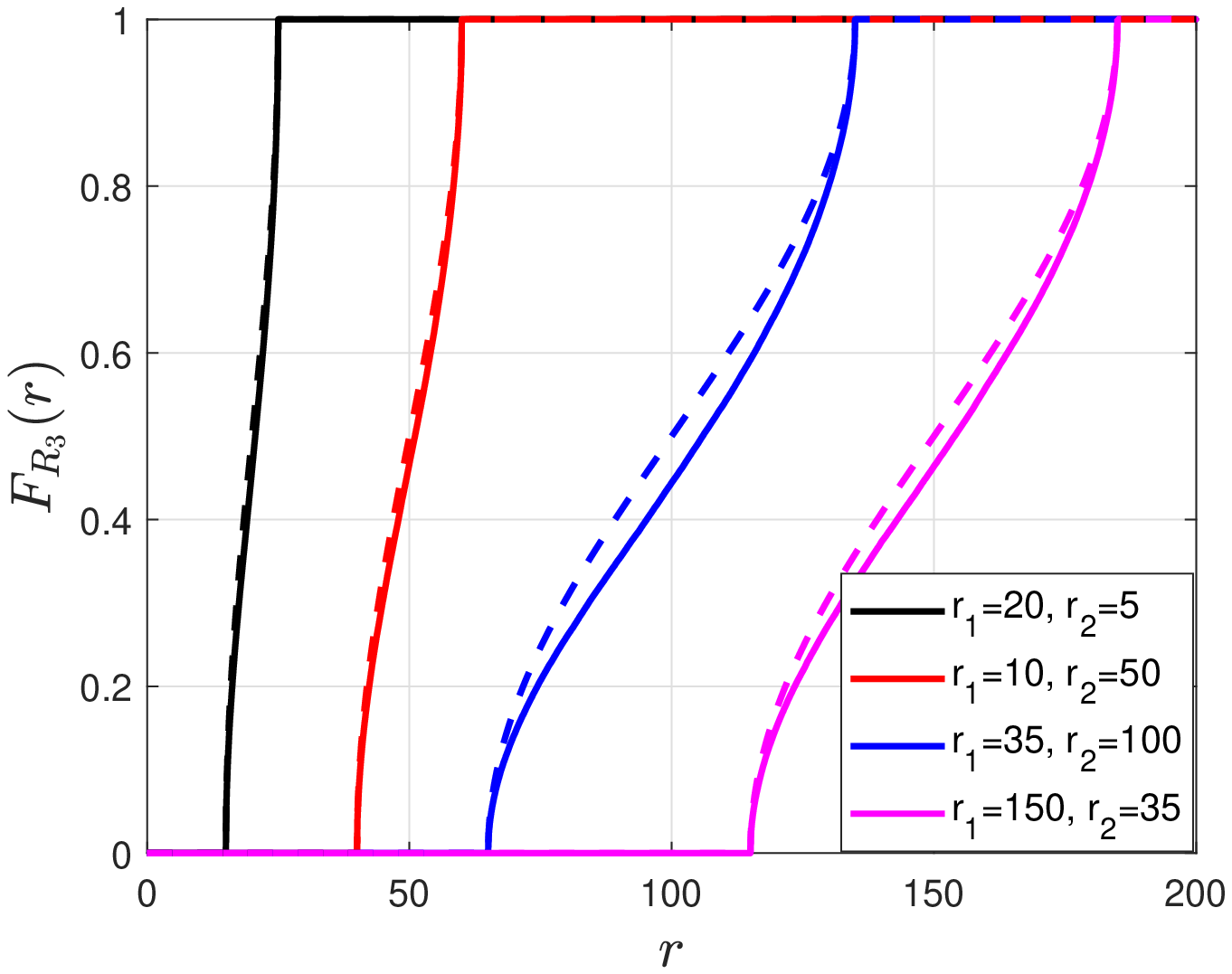}
\caption{The cdf of $R_3$ for different values of $r_1$ and $r_2$. Solid lines represent the simulations while dashed lines are used for the approximation in \eqref{F_R3}.}\label{R3_verify}
\end{minipage}\;\;
\begin{minipage}[htb]{0.47\linewidth}
\centering\includegraphics[width=3.1in]
{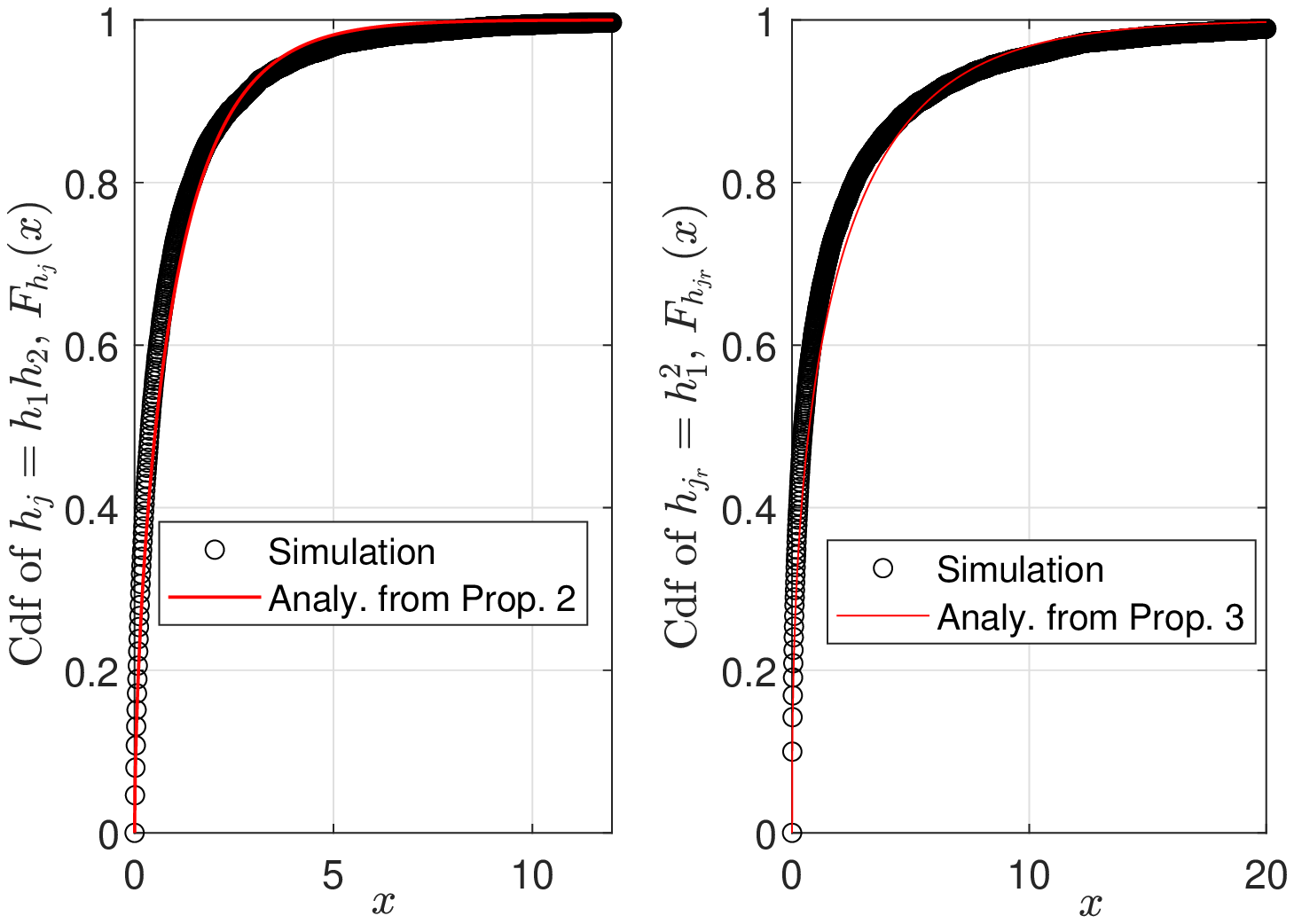}
\caption{The simulated and proposed analytical cdf of $h_j$ and $h_{j_r}$ in \eqref{hj_cdf} and \eqref{hjr_cdf}, respectively.}\label{cdf_hj}
\end{minipage}
\end{figure}

\textbf{\emph{Proposition 1:}} The distribution of the link distance $R_3$ is accurately approximated as
\begin{align}
&F_{R_3}(r)\!\approx\! \frac{1}{\pi}\! \cos^{-1} \left( \frac{- (r-|r_1-r_2|-\min(r_1,r_2))}{\min(r_1,r_2)} \right), \label{F_R3}
\end{align}
\begin{multline}
f_{R_3}(r)\approx \frac{1}{\pi \min(r_1,r_2)}\frac{1}{\sqrt{\left(1- \left(1+\frac{|r_1-r_2|-r}{\min(r_1,r_2)}\right)^2\right)}},  \;\;\;\; |r_1-r_2| \leq r \leq r_1+r_2. \label{f_R3}
\end{multline}
\textbf{\emph{Proof:}} From Fig. \ref{framework1} and \eqref{R3}, $|r_1-r_2| \leq R_3 \leq r_1+r_2$, {where the extreme values of $R_3$ are obtained from the scenarios where the BS, target and radar are on a straight line}. {Based on the uniform distribution of $\phi$ in $[ 0, 2\pi]$, we plot the exact distribution of $R_3$ in Fig. \ref{R3_verify} via simulations}. {The cdf is infinitely steep on both ends of the support and in the middle (at $\max(r_1, r_2)$) the value is close to 0.5 since the cdf is almost symmetric.} This distribution can thus be accurately modeled by taking the inverse of a scaled and translated cosine {function as follows}
\begin{align*}
&r=\min(r_1,r_2) \left( 1- \cos \left(\pi F_{R_3 }(r ) \right) \right) +|r_1-r_2|.
\end{align*}
By inverting this expression, we obtain the approximate cdf $F_{R_3 }(r)$ in \eqref{F_R3}. {From Fig. \ref{R3_verify} we observe that for different values of $r_1$ and $r_2$, the approximate analysis in \eqref{F_R3} closely matches the actual cdf of $R_3$ obtained via simulations, thus justifying the use of this approximation.} The corresponding pdf $f_{R_3 }(r)$ in \eqref{f_R3} is easily obtained by taking the derivative of \eqref{F_R3}. \qed

{\textbf{\emph{Remark 1:}} $F_{R_3}(r )$ remains unchanged if the values of $r_1$ and $r_2$ are interchanged as this is equivalent to the positions of the BS and radar in Fig. \ref{framework1} being swapped. Such an interchange does not impact the link distance $R_3$ and its statistics.}

{Due to the PPP nature of the interferers, the distance of the tBS to its nearest interferer, $\rho$, follows: 
\begin{align}
f_{\rho}(r)=2\pi \lambda r \exp(-\pi \lambda r^2), \;\;\;\; r \geq 0. \label{f_rho}
\end{align}}

The distance between the tRad and its nearest interferer, $\rho_r$ conditioned on $R_3$ follows
\begin{align}
f_{\rho_r \mid R_3}(r \mid R_3)=2\pi \lambda r \exp(-\pi \lambda (r^2 - R_3^2)), \;\;\;\; r \geq R_3,
\end{align}
due to the PPP distribution of interferers and the fact that the interferer lies greater than a distance $R_3$ away from the tRad\footnote{Since this work uses fixed distances $R_1=r_1$ and $R_2=r_2$ (to see the impact of varying these distances on radar-mode performance), we conditioned the distribution of $\rho_r$ on $R_3$ (a function of $r_1$ and $r_2$) for better accuracy of the statistics of $\rho_r$.}.

\subsubsection{Radar-only Network} In the radar-only network, we assume that the tTar lies a fixed distance $R_r=r_r$ from the tRad in a random direction. As the point process of interferers $\Phi_r$ is identical to $\Phi$, the distribution of the distance between the tRad in the radar-only network to its nearest interferer, $\rho_{\rm rad}$, is equivalent to the distribution in \eqref{f_rho}, i.e.,
\begin{align*}
f_{\rho_{\rm rad}}(r)=2\pi \lambda r \exp(-\pi \lambda r^2), \;\;\;\;  r \geq 0.
\end{align*}



%
%
%
%
%
%



\section{Proposed Strategy and Analysis Methodology}
\subsection{Dynamic Transmission Strategy (DTS) for ISAC}


In the radar-mode of ISAC, the signal received at the radar and BS (in bistatic and monostatic detection, respectively) is significantly weaker than the signal received by communication-mode users due to the double path loss associated with the signal in the radar-mode first going from the BS to the target and then from the target to the radar and BS. In contrast to the double path loss in the signal component of the radar-mode, the interference received at the radar and BS comes via direct links from interfering BSs; this further deteriorates the radar-mode's SINR. To combat this, we propose a {transmission strategy referred to as the DTS} in which BSs transmit with high power $P_{h}$ in one slot and with regular lower power $P_l$ in the remaining $M-1$ slots, where $M \geq 2$. The slot in which a BS transmits with higher power is chosen in an ALOHA fashion for different BSs and this is the slot in which the radar-mode is also active, i.e., ISAC takes place. The communication-mode, on the other hand, is active in all $M$ slots. For notational convenience, we refer to the slot in which the tBS transmits with $P_h$ as `slot 1'. Such a transmission strategy allows us to control the average network interference for both the communication and radar-modes as the probability of an interferer transmitting with $P_h$ is $1/M$. It also simultaneously gives the more vulnerable radar-mode, which is only active in slot 1, a higher transmission power to help improve its signal that undergoes a double path loss. Thus, the average radar signal power $P_h$ is higher than the average interference power, $P_{\rm avg}=(P_h + (M-1)P_l)/M$. 
%

This way we are able to improve the SINR of the radar-mode while having control on the impact it has on the communication-mode. {Note, however, that the price paid is reduced sensing as detection in the radar-mode only happens $1/M$ of the time. Since the communication requirements are generally much higher than sensing requirements and the radar-mode rides for free in the communication-mode's spectrum, the price paid is reasonable.} {It should also be emphasized that the DTS is a flexible strategy that allows control on the impact of both the quantity and quality of the radar-mode as well as on the quality of the communication-mode.} {Further, while the average signal power of the communication-mode is also $P_{\rm avg}$, the precise impact of this dynamic transmission strategy on the communication-mode is not obvious; we will thus also analyze it in this work.} We study the performance of both the communication-mode and radar-mode over $M$ time slots and assume that the tBS serves the tUE for $M$ time slots.

{It ought to be noted that while ISAC is promoted for more efficient use of the spectrum, which is a scarce resource, and reduced network power consumption (via fewer transmissions), the interference encountered by the nodes, including the communication-mode user, is higher with DTS than without DTS as $P_{\rm avg}>P_l$. Since the communication-mode typically pays for this in terms of resources (spectrum and transmit power) {and since the DTS can negatively impact the communication mode due to increased interference from the scenario without DTS}, protecting the communication-mode is a priority.}


 
\subsection{Methodology of Analysis}
Fixed-rate transmissions are used in this work, such that a message is sent with transmission rate $\log(1+\theta)$ corresponding to an SINR threshold of $\theta$. Such transmissions result in a throughput that is lower than the transmission rate because of possible outages. The performance of the communication-mode is typically measured using the coverage probability (i.e., the SINR ccdf), which is a measure of reliability, and throughput. Radar performance was traditionally analyzed using metrics like the detection probability and faulty error probability. More recently, however, works on ISAC such as \cite{jrc_irs1,jrc_sg5,Bazzi_JRC1}, coexisting communication and radar \cite{jrc_sg4} and even radar only \cite{sg_radar3,radar_scnrRef1} have switched to analyzing the performance of the radar-mode using the statistics of the SNR or SINR. In this work, we also focus on analyzing radar-mode performance using the statistics of the SINR at the radar and BS (in the bistatic and monostatic cases, respectively) as a measure of reliable detection.

\section{SINR Analysis}
\subsection{Preliminaries}
The fading coefficient between the tBS at \textbf{o} and the tUE at \textbf{c} (tTar at \textbf{t}) is $h_0$ ($h_1$), and between the tTar and the tRad at \textbf{r} is $h_2$. As the signal in the bistatic radar-mode goes from the tBS to the tTar and then to the tRad, the joint fading experienced at the tRad is $h_j=h_1 h_2$. On the other hand, in the monostatic radar-mode, the signal goes from the tBS to the tTar and then back to the tBS; the joint fading experienced at the tBS is $h_{j_r}=h_1^2$. Due to the Rayleigh fading assumption, $h_0$, $h_1$ and $h_2$ are independent unit mean exponential RVs.

While the exact statistics of the joint fading RVs $h_j$ and $h_{j_r}$ can be obtained, these approaches do not lead to tractability in obtaining the statistics of the SINR for the bistatic and monostatic radar-mode which we are interested in. We thus propose the use of tight approximations through curve fitting for the statistics of $h_j$ and $h_{j_r}$ that will allow tractability in obtaining the SINR statistics.

\textbf{\emph{Proposition 2:}} The cdf of the unit mean RV $h_j$ is approximated using 
\begin{align}
F_{h_j}(x) = \left(1-\exp \left(-\epsilon y \right)\right)^{m}, \label{hj_cdf}
\end{align}
where $m=\sqrt{7/20}$ and $\epsilon=\text{harmonic}(m)$.

\textbf{\emph{Proof:}} Based on the lower bound of the cdf of a gamma RV \cite{Alzer}, which is in a form that allows tractability in obtaining the statistics of the SINR for a PPP of interferers, using the fact that $\mathbb{E}[h_j]=\mathbb{E}[h_1]\mathbb{E}[h_2]=1$  and via fitting we obtain the expression in \eqref{hj_cdf}. In Fig. \ref{cdf_hj} we observe that the exact distribution based on simulations closely matches the analytical approximate in \eqref{hj_cdf}. \qed

\textbf{\emph{Proposition 3:}} The cdf of the RV $h_{j_r}$ with $\mathbb{E}[h_{j_r}]=2$ is approximated using 
\begin{align}
F_{h_{j_r}}(x) = \left(1-\exp \left(-\epsilon_r y \right)\right)^{m_r}, \label{hjr_cdf}
\end{align}
where $m_r=\sqrt{3/20}$ and $\epsilon_r=\text{harmonic}(m)/2$. 

\textbf{\emph{Proof:}} Along the lines of {the proof of} Proposition 2 and using the fact that $\mathbb{E}[h_{j_r}]=\mathbb{E}[{h_1^2}]=\text{Var}[h_1]+(\mathbb{E}[h_1])^2=2$ as $h_1\sim \exp(1)$, we obtain \eqref{hjr_cdf}. In Fig. \ref{cdf_hj} we observe that the exact distribution based on simulations closely matches the analytical approximate in \eqref{hjr_cdf}. \qed


The intercell interference experienced at a receiver located at $\textbf{z}$ scaled to unit transmission power is of the form 
\begin{align}
I_{z}=\sum_{\textbf{x} \in \Phi}  g_{\textbf{x}_z} {\|\textbf{x}-\textbf{z}\|}^{-\eta}, \label{interf}
\end{align} 
where $z \in \{c, r,o \}$ denotes the interference at $\textbf{c}$, $\textbf{r}$ and $\textbf{o}$ in the cellular mode, bistatic radar-mode and monostatic radar-mode, respectively. Here, $g_{\textbf{x}_z}$ denotes the fading coefficient from the interfering BS at $\textbf{x}$ to the receiver at $\textbf{z}$. 

For the case of $z \in \{ r,o \}$, i.e., the bistatic and monostatic radar modes, let $\psi$ denote the guard zone distance (i.e., the distance between the receiver and the nearest interferer), where $\psi=\{\rho_r, \rho \}$ for the bistatic and monostatic radar-modes, respectively. The intercell interference for $z \in \{ r,o \}$ scaled to unit transmission power at $\textbf{z}$ can be rewritten as 
\begin{align}
I_{z}=\sum_{\substack{\textbf{x}\in\Phi\\ \|\textbf{x}-\textbf{z}\|> \psi }}     g_{\textbf{x}_z} {\|\textbf{x}-\textbf{z}\|}^{-\eta} + \sum_{\substack{\textbf{x}\in\Phi\\ \|\textbf{x}-\textbf{z}\|= \psi }}     g_{\textbf{x}_z} {\|\textbf{x}-\textbf{z}\|}^{-\eta}, \label{interfCond}
\end{align} 

The LT of $I_z$ for $z \in \{ r,o \}$ conditioned on ${\psi}$ is given by
{ \begin{align}
\mathcal{L}_{I_z \mid \psi}(s)&=\exp \left({   \frac{-2 \pi \lambda s }{(\eta-2){{\psi}}^{\eta-2}} { }_2F_1  \left( 1,1  -  \delta; 2 - \delta; \frac{-s}{{{\psi}}^{\eta}} \right)   }\right)  \frac{1}{1+s \psi^{-\eta}}  \label{L_I_cond}\\
&\stackrel{\eta=4}= \exp \left(-\pi \lambda \sqrt{s} \tan^{-1} \left({\sqrt{s}}{{\psi}^{-2}} \right) \right) \frac{1}{1+s \psi^{-\eta}} .
\end{align} }

We obtain the first term in \eqref{L_I_cond} using the probability generating functional (pgfl) of the PPP, the guard zone distance ${\psi}$ and the fact that $g_{\textbf{x}_z} \sim \exp(1)$ due to the Rayleigh fading assumption \cite{mhaenggi_Book}. The second term in \eqref{L_I_cond} comes from $g_{\textbf{x}_z} \sim \exp(1)$ and due to the fact that an interferer is conditioned to exist at a distance $\psi$ from the receiver according to {the second term in} \eqref{interfCond}.

The LT of $I_c$, the intercell interference in the communication-mode at the tUE located at $\textbf{c}$ scaled to unit transmission power, conditioned on the link distance $R_0$ is
\begin{align}
\mathcal{L}_{I_c \mid R_0}(s)&=\exp \left({   \frac{-2 \pi \lambda s }{(\eta-2){R_0}^{\eta-2}} { }_2F_1  \left( 1,1  -  \delta; 2 - \delta; \frac{-s}{{R_0}^{\eta}} \right)   }\right)     \label{L_I}\\
&\stackrel{\eta=4}= \exp \left(-\pi \lambda \sqrt{s} \tan^{-1} \left({\sqrt{s}}{R_0^{-2}} \right) \right) 
\end{align}
Similar to \eqref{L_I_cond}, we obtain \eqref{L_I} using the pgfl of the PPP, the guard zone distance $R_0$ and the fact that $g_{\textbf{x}_z} \sim \exp(1)$.

\subsection{SINR Analysis of ISAC Network employing the DTS}
The SINR of the communication-mode at the tUE in an ISAC network employing the DTS when the tBS transmits with power $P_{\chi}$, $\chi \in \{l,h\}$ is
\begin{align}
&{\rm SINR_c^{\chi}}= \frac{ P_{\chi} h_0 R_0^{-\eta}  }{ P_{\rm avg }I_c +  \sigma^2 } , \label{SINR_c}
\end{align} 
{where $P_{\rm avg }I_c$ is} the interference experienced at the tUE located at $\textbf{c}$.

\textbf{\emph{Lemma 1:}} The ccdf of the communication-mode SINR when the tBS transmits with power $P_{\chi}$, $\chi \in \{l,h\}$ in an ISAC network employing DTS is
\begin{align}
&\mathbb{P}({\rm SINR_c^{\chi}}>\theta ) =  \mathbb{E}_{R_0} \left[   \mathcal{L}_{I_c \mid R_{0}}  \left( \frac{ \theta R_0^{\eta} P_{\rm avg}}{P_{\chi} }  \right) \exp\left( \frac{- \theta R_0^{\eta} \sigma^2}{P_{\chi} }  \right)  \right], \label{ccdfSINR_c}
\end{align}
{where $\mathcal{L}_{I_c \mid R_{0}}(s)$ is given in \eqref{L_I}.} 

\textbf{\emph{Proof:}} Rewriting the SINR expression in \eqref{SINR_c} we obtain
\begin{align*}
\mathbb{P}({\rm SINR_c^{\chi}}>\theta )=\mathbb{E}\left[ \bar{F}_{h_0} \left( \frac{\theta R_0^{\eta} }{ P_{\chi}} \left( P_{\rm avg} I_c + \sigma^2 \right) \right) \right].
\end{align*}
Since $h_0 \sim \exp(1)$, using the definition of the LT, we obtain \eqref{ccdfSINR_c}. \qed

{\textbf{\emph{Theorem 1:}} The average ccdf of the communication-mode SINR in an ISAC network employing DTS is 
\begin{multline}
\mathbb{P}({\rm SINR_c^{\rm avg}}>\theta )=    \mathbb{E}_{R_0} \Bigg[ \frac{1}{M}  \mathcal{L}_{I_c \mid R_{0}}  \left( \frac{ \theta R_0^{\eta} P_{\rm avg}}{P_h }  \right) \exp\left( \frac{- \theta R_0^{\eta} \sigma^2}{P_h }  \right)  \\
 + \frac{M-1}{M}  \mathcal{L}_{I_c \mid R_{0}}  \left( \frac{ \theta R_0^{\eta} P_{\rm avg}}{P_l }  \right) \exp\left( \frac{- \theta R_0^{\eta} \sigma^2}{P_l }  \right)  \Bigg]. \label{thm1_2}  
\end{multline} }
{\textbf{\emph{Proof:}} Since with the DTS the BS transmits with $P_h$ in 1 of $M$ slots and with $P_l$ in the remaining $M-1$ slots, by weighted averaging over $M$ slots, we obtain 
\begin{align*}
&\mathbb{P}({\rm SINR_c^{\rm avg}}>\theta )= \frac{1}{M} \mathbb{P}({\rm SINR_c^{h}}>\theta ) + \frac{M-1}{M} \mathbb{P}({\rm SINR_c^{l}}>\theta ) .
\end{align*}
Applying Lemma 1, \eqref{thm1_2} is obtained. \qed   }

The SINR of the bistatic radar-mode at the tRad in an ISAC network employing the DTS is
\begin{align}
&{\rm SINR_{r,b}}= \frac{ P_h h_1 h_2 r_1^{-\eta} r_2^{-\eta} }{ P_{\rm avg } I_r+  \sigma^2 } = \frac{ P_h h_j r_1^{-\eta} r_2^{-\eta} }{ P_{\rm avg } I_r+  \sigma^2 } , \label{SINR_r_bi}
\end{align} 
where $P_{\rm avg} I_r$ is the interference experienced at the tRad located at $\textbf{r}$.

\textbf{\emph{Lemma 2:}} The ccdf of the bistatic radar-mode's SINR in an ISAC network employing DTS is
\begin{align}
&\mathbb{P}({\rm SINR_{r,b}}>\theta )=  1 \!-\!  \mathbb{E}_{R_3,\rho_r} \left[  \left(  1\! - \! \mathcal{L}_{I_r \mid \rho_r}  \left( \frac{\epsilon  \theta r_1^{\eta} r_2^{\eta} P_{\rm avg} }{P_h}  \right) \! \exp \left( \frac{-\epsilon  \theta r_1^{\eta} r_2^{\eta} \sigma^2 }{P_h}  \right)  \right)^{m}  \right], \label{ccdfSINR_r_bi}
\end{align} 
{where $\mathcal{L}_{I_r \mid \rho_r}(s)$ is obtained from \eqref{L_I_cond}.}

\textbf{\emph{Proof:}} Rewriting the SINR expression in \eqref{SINR_r_bi} we obtain
\begin{align*}
&\mathbb{P}({\rm SINR_{r,b}}>\theta )=\mathbb{E}\left[ \bar{F}_{h_j} \left( \frac{\theta r_1^{\eta} r_2^{\eta} }{ P_h} \left(P_{\rm avg }I_r + \sigma^2 \right) \right) \right].
\end{align*} 
Using the cdf of $h_j$ in \eqref{hj_cdf} along with the definition of the LT, we obtain \eqref{ccdfSINR_r_bi}.  \qed

The SINR of the monostatic radar-mode at the tBS in an ISAC network employing the DTS is
\begin{align}
&{\rm SINR_{r,m}}= \frac{ P_h h_1 h_1 r_1^{-2\eta}  }{ P_{\rm avg } I_o+  \sigma^2 } = \frac{ P_h h_{j_r} r_1^{-2\eta}  }{ P_{\rm avg } I_o+  \sigma^2 } , \label{SINR_r_mo}
\end{align} 
where $P_{\rm avg} I_{o}$ is the interference experienced at the tBS located at $\textbf{o}$.

\textbf{\emph{Lemma 3:}} The ccdf of the monostatic radar-mode's SINR in an ISAC network employing DTS is
\begin{align}
&\mathbb{P}({\rm SINR_{r,m}}>\theta )=  1 \!-\!  \mathbb{E}_{\rho} \left[  \left(  1 \!-\! \mathcal{L}_{I_o \mid \rho}  \left( \frac{\epsilon_r  \theta r_1^{\eta} r_2^{\eta} P_{\rm avg} }{P_h}  \right) \exp \left( \frac{-\epsilon_r  \theta r_1^{\eta} r_2^{\eta} \sigma^2 }{P_h}  \right)  \right)^{m_r}  \right], \label{ccdfSINR_r_mo}
\end{align} 
{where $\mathcal{L}_{I_o \mid \rho}(s)$ is obtained from \eqref{L_I_cond}.}

\textbf{\emph{Proof:}} Rewriting the SINR expression in \eqref{SINR_r_mo} we obtain
\begin{align*}
\mathbb{P}({\rm SINR_{r,m}}>\theta )=\mathbb{E}\left[ \bar{F}_{h_{j_r}} \left( \frac{\theta r_1^{2\eta}  }{ P_h} \left(P_{\rm avg }I_o + \sigma^2 \right) \right) \right].
\end{align*}
Using the cdf of $h_{j_r}$ in \eqref{hjr_cdf} along with the definition of the LT, we obtain \eqref{ccdfSINR_r_mo}.  \qed

{\textbf{\emph{Theorem 2:}} The joint SINR ccdf of the radar-mode in an ISAC network employing DTS, where there is cooperation between bistatic and monostatic detection is 
\begin{align}
&\mathbb{P}({\rm SINR_{r,joint}}>\theta ) = \mathbb{P}({\rm SINR_{r,b}}>\theta )+\mathbb{P}({\rm SINR_{r,m}}>\theta ) - \mathbb{P}({\rm SINR_{r,b}}>\theta )\mathbb{P}({\rm SINR_{r,m}}>\theta ). \label{ccdfSINR_r_joint}
\end{align} }
{\textbf{\emph{Proof:}} By definition,
\begin{align*}
&\mathbb{P} ({\rm SINR_{r,joint}}>\theta )= \mathbb{P} \left( \{ {\rm SINR_{r,b}}>\theta \} \bigcup \{ {\rm SINR_{r,m}}>\theta \} \right).
\end{align*}
 Since using the exact statistics of $h_1 h_2$ and $h_1^2$ do not result in tractability in computing the SINR statistics, we proposed using $h_j$ and and $h_{j_r}$, respectively in Propositions 2 and 3, which are independent. Based on these approximations, the two detection events in the equation above are independent and \eqref{ccdfSINR_r_joint} follows from the probability of the union of these two events.} \qed

\subsection{SINR Analysis of ISAC Network without DTS}
Since the proposed DTS uses the higher $P_h$ in one slot, the normal transmission power in the network without DTS is $P_l$.

The SINR of the communication-mode at the tUE in an ISAC network without the DTS is
\begin{align}
&{\rm SINR_c^{\text{no DTS}}}= \frac{ P_l h_0 R_0^{-\eta}  }{ P_l I_c+  \sigma^2 } . \label{SINR_c_noDTS}
\end{align} 

\textbf{\emph{Lemma 4:}} The ccdf of the communication-mode SINR in an ISAC network without DTS is
\begin{align}
&\mathbb{P}({\rm SINR_c^{\text{no DTS}}}>\theta )=  \mathbb{E}_{R_0} \left[   \mathcal{L}_{I_c \mid R_{0}}  \left( \theta R_0^{\eta}  \right) \exp\left( \frac{- \theta R_0^{\eta} \sigma^2}{P_l }  \right)  \right], \label{ccdfSINR_c_noDTS}
\end{align} 
{where $\mathcal{L}_{I_c \mid R_{0}}(s)$ is given in \eqref{L_I}.}

\textbf{\emph{Proof:}} Using \eqref{SINR_c_noDTS} we obtain
\begin{align*}
&\mathbb{P}({\rm SINR_c^{\text{no DTS}}}>\theta )=\mathbb{E}\left[ \bar{F}_{h_0} \left( \frac{\theta R_0^{\eta} }{ P_{l}} \left( P_l I_c + \sigma^2 \right) \right) \right].
\end{align*}
 Since $h_0 \sim \exp(1)$, using the definition of the LT, we obtain \eqref{ccdfSINR_c_noDTS}.

\textbf{\emph{Remark 2:}} The SINR of the communication-mode in a network without ISAC, i.e., a traditional cellular network, is identical to the SINR of the communication-mode in an ISAC network without DTS as the communication-mode is not impacted by ISAC when DTS is not used.

The SINR of the bistatic radar-mode at the tRad in an ISAC network without the DTS is
\begin{align}
&{\rm SINR_{r,b}^{\text{no DTS}}}= \frac{ P_l h_j r_1^{-\eta} r_2^{-\eta} }{ P_l I_r+  \sigma^2 } . \label{SINR_r_bi_noDTS}
\end{align} 

\textbf{\emph{Lemma 5:}} The ccdf of the bistatic radar-mode's SINR in an ISAC network without DTS is
\begin{align}
&\mathbb{P}({\rm SINR_{r,b}^{\text{no DTS}}}>\theta )=  1 -  \mathbb{E}_{R_3,\rho_r} \left[  \left(  1 - \mathcal{L}_{I_r \mid \rho_r}  \left( \epsilon  \theta r_1^{\eta} r_2^{\eta}  \right) \exp \left( \frac{-\epsilon  \theta r_1^{\eta} r_2^{\eta} \sigma^2 }{P_l}  \right)  \right)^{m}  \right], \label{ccdfSINR_r_bi_noDTS}
\end{align} 
{where $\mathcal{L}_{I_r \mid \rho_r}(s)$ is obtained from \eqref{L_I_cond}.}

\textbf{\emph{Proof:}} Using \eqref{SINR_r_bi_noDTS} we obtain
\begin{align*}
\mathbb{P}({\rm SINR_{r,b}^{\text{no DTS}}}>\theta )=\mathbb{E}\left[ \bar{F}_{h_j} \left( \frac{\theta r_1^{\eta} r_2^{\eta} }{ P_l} \left(P_l I_r + \sigma^2 \right) \right) \right].
\end{align*}
 Using the cdf of $h_j$ in \eqref{hj_cdf} along with the definition of the LT, we obtain \eqref{ccdfSINR_r_bi_noDTS}. \qed

The SINR of the monostatic radar-mode at the tBS in an ISAC network without the DTS is
\begin{align}
&{\rm SINR_{r,m}^{\text{no DTS}}}= \frac{ P_l h_{j_r} r_1^{-2\eta}  }{ P_l I_o+  \sigma^2 } . \label{SINR_r_mo_noDTS}
\end{align} 

\textbf{\emph{Lemma 6:}} The ccdf of the monostatic radar-mode's SINR in an ISAC network without DTS is
\begin{align}
&\mathbb{P}({\rm SINR_{r,m}^{\text{no DTS}}}>\theta )=  1 -  \mathbb{E}_{\rho} \left[  \left(  1 - \mathcal{L}_{I_o \mid \rho}  \left( \epsilon_r  \theta r_1^{2\eta}   \right) \exp \left( \frac{-\epsilon_r  \theta r_1^{2\eta}  \sigma^2 }{P_l}  \right)  \right)^{m_r}  \right], \label{ccdfSINR_r_mo_noDTS}
\end{align} 
{where $\mathcal{L}_{I_o \mid \rho}(s)$ is obtained from \eqref{L_I_cond}.}

\textbf{\emph{Proof:}} Using \eqref{SINR_r_mo_noDTS} we obtain
\begin{align*}
\mathbb{P}({\rm SINR_{r,m}^{\text{no DTS}}}>\theta )=\mathbb{E}\left[ \bar{F}_{h_{j_r}} \left( \frac{\theta r_1^{2\eta}  }{ P_l} \left(P_l I_o + \sigma^2 \right) \right) \right].
\end{align*}
Using the cdf of $h_{j_r}$ in \eqref{hjr_cdf} along with the definition of the LT, we obtain \eqref{ccdfSINR_r_mo_noDTS}. \qed

{\textbf{\emph{Theorem 3:}} The joint SINR ccdf of the radar-mode in an ISAC network without DTS, where there is cooperation between bistatic and monostatic detection is approximated as
\begin{align}
\mathbb{P}({\rm SINR_{r,joint}^{\text{no DTS}}}>\theta )&= \mathbb{P}({\rm SINR_{r,b}^{\text{no DTS}}}>\theta )+\mathbb{P}({\rm SINR_{r,m}^{\text{no DTS}}}>\theta ) -  \nonumber \\ 
& \mathbb{P}({\rm SINR_{r,b}^{\text{no DTS}}}>\theta )\mathbb{P}({\rm SINR_{r,m}^{\text{no DTS}}}>\theta ). \label{ccdfSINR_r_joint_noDTS}
\end{align}
\textbf{\emph{Proof:}} By definition, 
\begin{align*}
\mathbb{P}({\rm SINR_{r,joint}^{\text{no DTS}}}>\theta )&= \mathbb{P}\left( \{{\rm SINR_{r,b}^{\text{no DTS}}}>\theta \} \bigcup \{ {\rm SINR_{r,m}^{\text{no DTS}}}>\theta \} \right).
\end{align*}
Along the lines of the proof of Theorem 2, the two detection events in the equation above are independent and \eqref{ccdfSINR_r_joint_noDTS} follows from the probability of the union of these two events. \qed }

\subsection{SINR Analysis of Radar-only Network}
The SINR of the radar-mode at the tRad in a radar-only network via monostatic detection is
\begin{align}
&{\rm SINR}^{\text{Rad-only}}= \frac{ P_r h_r^2 R_r^{-2\eta} }{ P_r I_{r_r}+  \sigma^2 } = \frac{ P_r h_{j_r} R_r^{-2\eta} }{ P_r I_{r_r}+  \sigma^2 } , \label{SINR_radOnly}
\end{align} 
where $h_r$ is fading coefficient from the tRad at \textbf{o} to the tTar and the intercell interference at the tRad is $P_r I_{r_r}$. Here, $I_{r_r}=\sum_{\textbf{x} \in \Phi_r}  g_{\textbf{x}} {\|\textbf{x}\|}^{-\eta}$; since $\Phi_r$ like $\Phi$ has intensity $\lambda$, the LT of $I_{r_r}$ is conditioned on $\rho_{\rm rad}$ and can be obtained from \eqref{L_I_cond} as $\mathcal{L}_{I_{r_r} \mid \rho_{\rm rad}}(s)$. Since $h_r \sim \exp(1)$, $h_r^2 \equiv h_{j_r}$ in \eqref{SINR_radOnly} and we use the statistics of $h_{j_r}$ in \eqref{hjr_cdf} from Proposition 3.

%
%


\textbf{\emph{Lemma 7:}} The ccdf of the SINR at the tRad in a radar-only network is
\begin{align}
&\mathbb{P}({\rm SINR_r^{\text{Rad-only}}}>\theta )=  1 -  \mathbb{E}_{\rho_{\rm rad}} \left[  \left(  1 - \mathcal{L}_{I_{r_r} \mid \rho_{\rm rad}}  \left( \epsilon_r  \theta R_r^{2\eta}   \right) \exp \left( \frac{-\epsilon_r  \theta R_r^{2\eta}  \sigma^2 }{P_r}  \right)  \right)^{m_r}  \right], \label{ccdfSINR_radOnly}
\end{align} 
{where $\mathcal{L}_{I_{r_r} \mid \rho_{\rm rad}}(s)$ is obtained from \eqref{L_I_cond}.}

\textbf{\emph{Proof:}} Rewriting the SINR expression in \eqref{SINR_radOnly} we obtain 
\begin{align*}
\mathbb{P}({\rm SINR_r^{\text{Rad-only}}}>\theta )=\mathbb{E}\left[ \bar{F}_{h_{j_r}} \left( \frac{\theta R_r^{2\eta} }{ P_r} \left(P_r I_{r_r} + \sigma^2 \right) \right) \right].
\end{align*}
Using the cdf of $h_{j_r}$ in \eqref{hjr_cdf} and the definition of the LT, we obtain \eqref{ccdfSINR_radOnly}. \qed

\section{Results}
\begin{figure}
\begin{minipage}[htb]{0.49\linewidth}
\centering\includegraphics[width=\linewidth]
{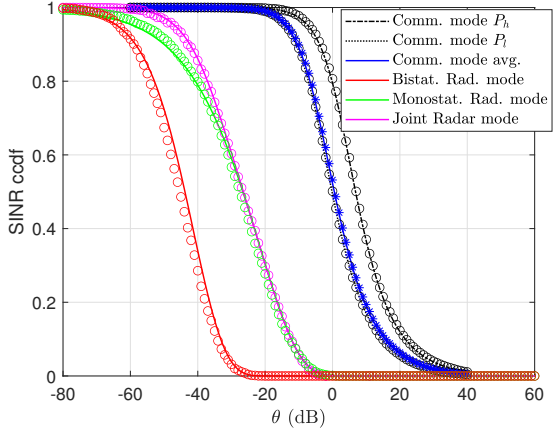}
\subcaption{The ISAC network employing DTS.\\$\;$}\label{verify1}
\end{minipage}\;\;\;
\begin{minipage}[htb]{0.49\linewidth}
\centering\includegraphics[width=\linewidth]
{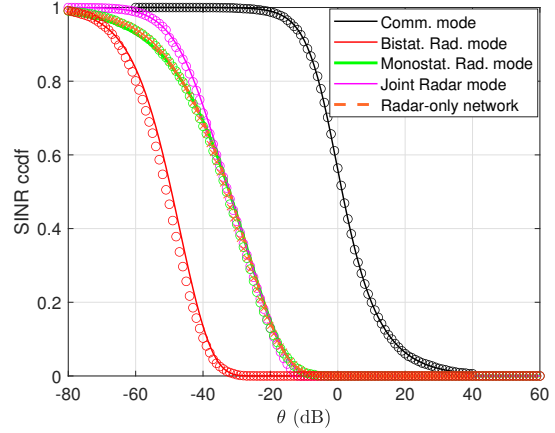}
\subcaption{The ISAC network without DTS (solid lines) and the radar-only network (dashed lines).}\label{verify2}
\end{minipage}
\begin{minipage}[htb]{0.49\linewidth}
\centering\includegraphics[width=\linewidth]
{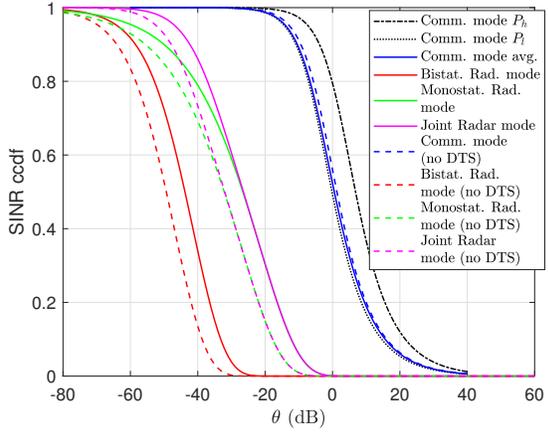}
\subcaption{The ISAC network with and without DTS (analytical only).}\label{verify3}
\end{minipage}
\caption{The ccdf of the SIR using $r_1=5v$ and $r_2=15v$. Lines (markers) represent the analytical expressions (simulations). DTS is employed using $P_h=5$ and $M=10$.}\label{verify}
\end{figure}

In this section, unless mentioned otherwise, we consider BS intensity $\lambda=10^{-5}$, $\eta=4$, $P_l=1$ and $v=1/(60\sqrt{\lambda}) $. Simulations are repeated $10^4$ times. {In the ISAC network with DTS, the radar-mode is only active in slot 1 where the transmit-power is $P_h$ and the performance is plotted accordingly.} For fairness of comparison we also set $R_r=r_1$ and $P_r=P_l=1$. Since $R_r=r_1$ and $P_r=P_l$, in this section, the performance of the radar-only network (from \eqref{ccdfSINR_radOnly}) is identical to the performance of monostatic detection in the ISAC network without DTS (i.e., from \eqref{ccdfSINR_r_mo_noDTS}). This will be verified in Fig. \ref{verify} too.



In Fig. \ref{verify} we plot the ccdf of the SINRs of interest in the ISAC network employing the DTS, the ISAC network without  DTS and the radar-only network. Figs. \ref{verify1} and \ref{verify2} verify the accuracy of our mathematical analysis as the analytical expressions match the simulations well. As anticipated, the radar-mode in the case of both bistatic and monostatic detection always performs worse than its communication counterpart due to the impact of the double path-loss component; in fact, even with joint detection, the performance of the radar-mode is worse than the communication-mode highlighting the significant deterioration caused by the double path-loss. In Fig. \ref{verify2} we verify that the performance of the radar-mode with monostatic detection in the ISAC network without DTS is identical to that in the radar-only network. This is because $P_r=P_l$ and $R_r=r_1$ in the plotted figure for fairness of comparison; however, note that in ISAC network, the radar-mode rides in the communication-mode's spectrum for free and does not require its own transmit power as it uses the cellular transmit signal. In the radar-only network, on the other hand, while we obtain the same performance as the monostatic ISAC without DTS, the radar-mode pays for its own spectrum and transmission power. Further, with the proposed DTS solution as well as with joint detection using the monostatic and bistatic detection together, we are able to improve the performance of radar detection in the ISAC network significantly more than the radar-only network. The improvement with using bistatic detection jointly with the monostatic detection also stems from the fact that the radar-only network (which is monostatic) suffers a double path-loss of the form $R_r^{-2\eta}$ while in the ISAC network with bistatic radar detection the path loss is of the form $R_1^{-\eta}r_2^{-\eta}$. By increasing the intensity of listening radars, which are low cost and passive, this approach in the ISAC network has the potential to control the deterioration caused by the double path loss term. This highlights the superiority of ISAC over a radar-only network not just in terms of performance improvement via DTS and joint detection but also in terms of reduction of expense for both spectrum and power consumption.

In Fig. \ref{verify3} we observe that employing DTS improves the performance of the radar-mode; this is seen for the bistatic detection, the monostatic detection as well as for joint detection. The impact of employing DTS on the communication-mode is more complicated due to the trade off between improved average performance due to the higher power $P_h$ in one slot and the increased interference for the remaining slots. We observe that the average communication-mode performance when DTS is employed for the chosen parameters is worse than the performance without DTS by less than 1 dB. On the other hand, the improvement with DTS in the radar-mode is up to 5.5 dB. This highlights that with careful selection of $P_h$ and $M$, DTS can result in significantly higher gains for the radar-mode than the deterioration of the communication-mode.


\begin{figure}
\begin{minipage}[htb]{0.47\linewidth}
\centering\includegraphics[width=\linewidth]
{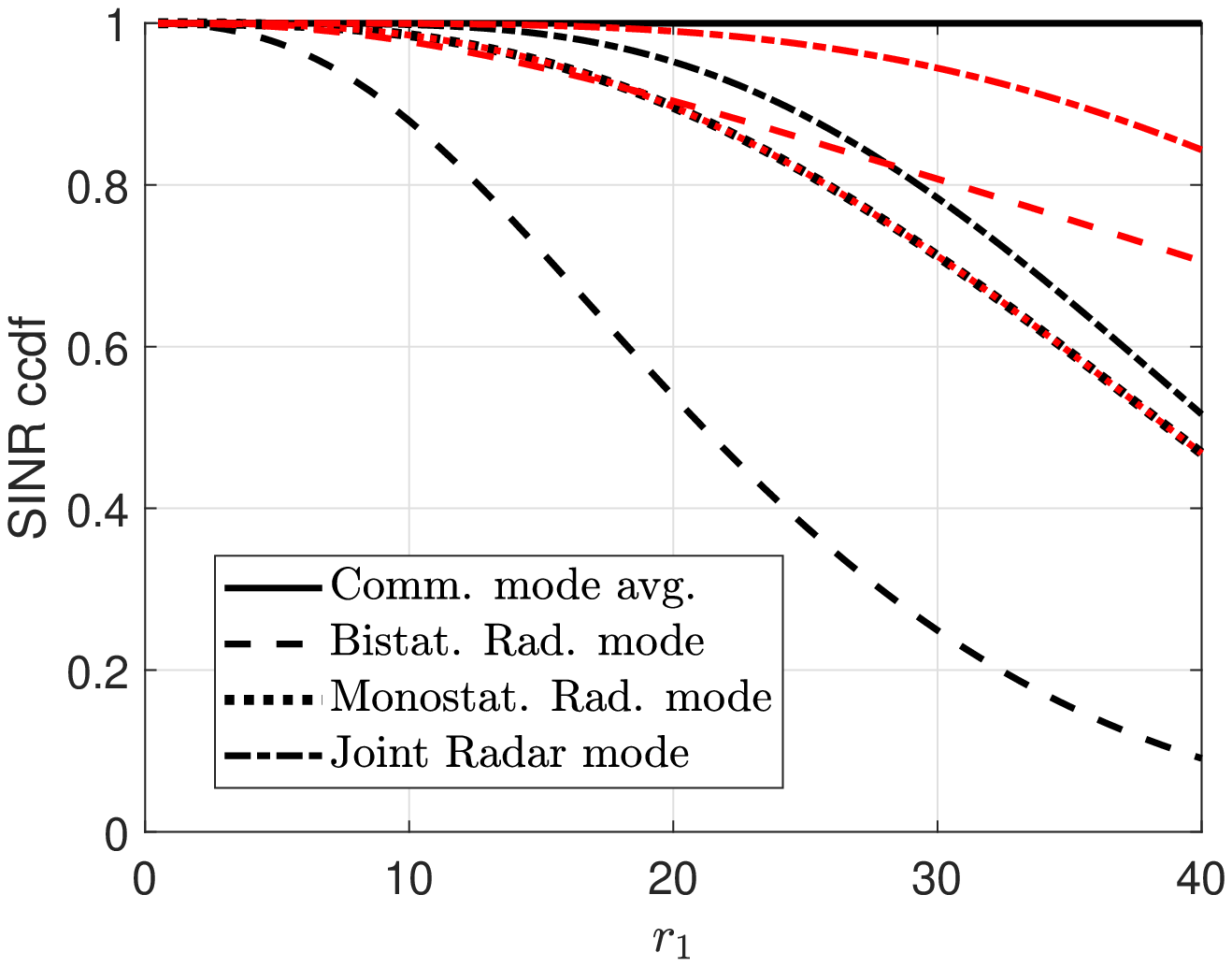}
\subcaption{With DTS}\label{vsR1DTS}
\end{minipage}\;\;\;
\begin{minipage}[htb]{0.47\linewidth}
\centering\includegraphics[width=\linewidth]
{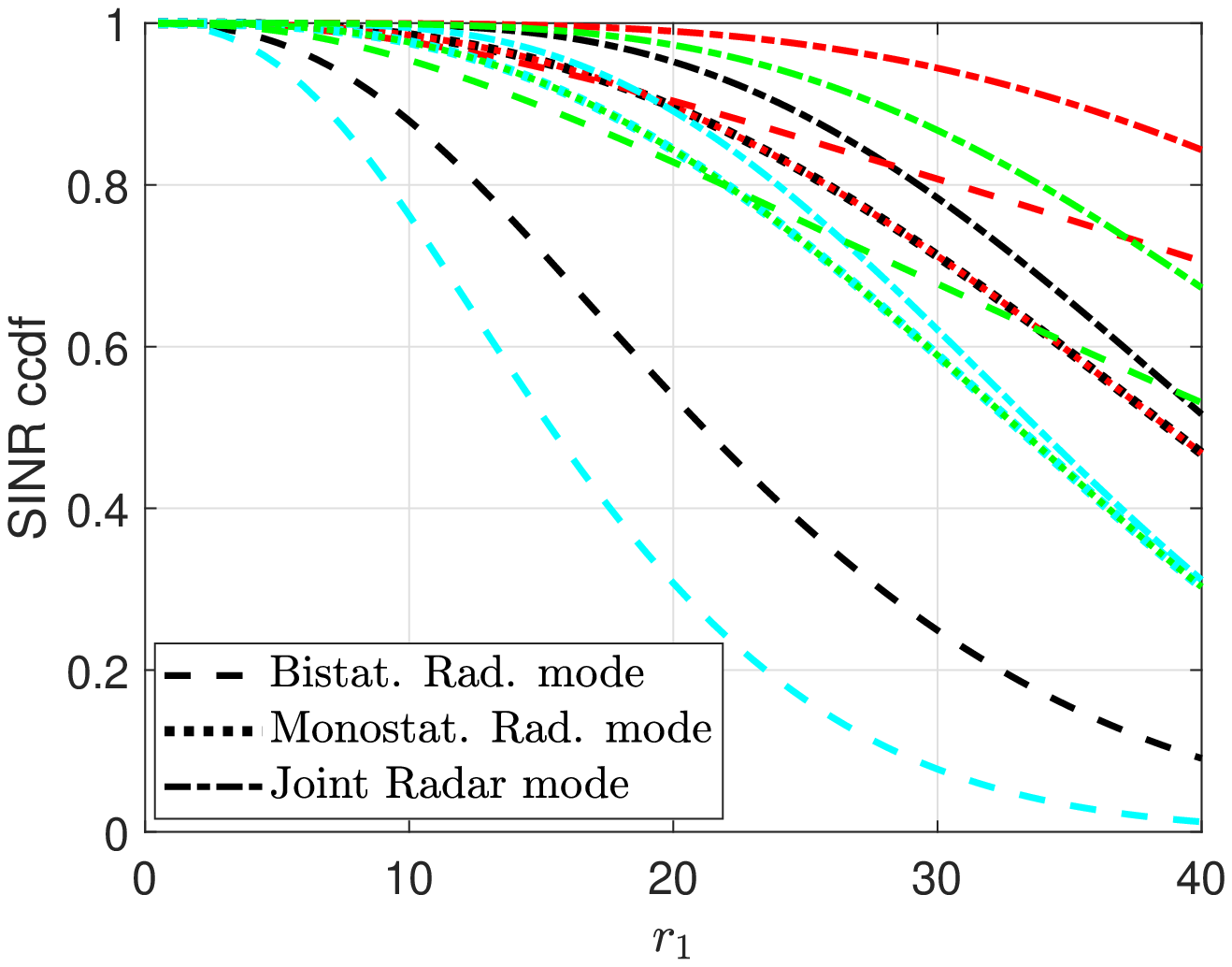}
\subcaption{With and without DTS}\label{vsR1DTSandNoDTS}
\end{minipage}
\caption{Performance of the ISAC network as $r_1$ is increased using $\theta=-40$ dB. Black (red) lines represent $r_2=15v$ ($r_2=5v$) with DTS using $P_h=5$ and $M=10$. Cyan (green) lines represent $r_2=15v$ ($r_2=5v$) without DTS.}\label{vsR1}
\end{figure}

Fig. \ref{vsR1} is a plot of the SINR ccdfs in an ISAC network as $r_1$ is increased. We plot these with and without DTS for the radar-mode for bistatic, monostatic and joint detection as well as the average for the communication mode. In Fig. \ref{vsR1DTS}, we observe that, as anticipated, the SINR ccdf of the communication-mode is not impacted by $r_1$, while the reliability of detection in each of bistatic, monostatic and joint detection decrease with increasing $r_1$ as the path loss decreases. Further, we observe that the impact of increasing $r_1$ on monostatic detection is stronger than on bistatic detection when $r_2$ is small as the SINR of monostatic detection falls as $r_1^{-2\eta}$ while the SINR in bistatic detection only falls as $r_1^{-\eta}$. This holds true on average in general, however, when $r_2$ is large, due to the chosen parameters, the impact of increasing $r_1$ is amplified more in this case and thus for the larger $r_2$ in Fig. \ref{vsR1DTS} we observe that the bistatic mode is impacted more by increasing $r_1$. As $r_2$ is decreased, the performance of bistatic detection increases while the performance of monostatic detection remains unchanged; as a result, an increase in the performance of joint detection is also observed. In Fig. \ref{vsR1DTSandNoDTS}, the same trends are observed for the case without DTS. As anticipated, the performance of the radar-mode with DTS is always superior to its counterpart without DTS. We also note that the amount of increase in performance by incorporating DTS is roughly the same for both bistatic and monostatic detection as well as joint detection. This highlights the significance of using DTS to improve radar-mode performance even if only one detection technique is available.

\begin{figure}
\begin{minipage}[htb]{0.47\linewidth}
\centering\includegraphics[width=\linewidth]
{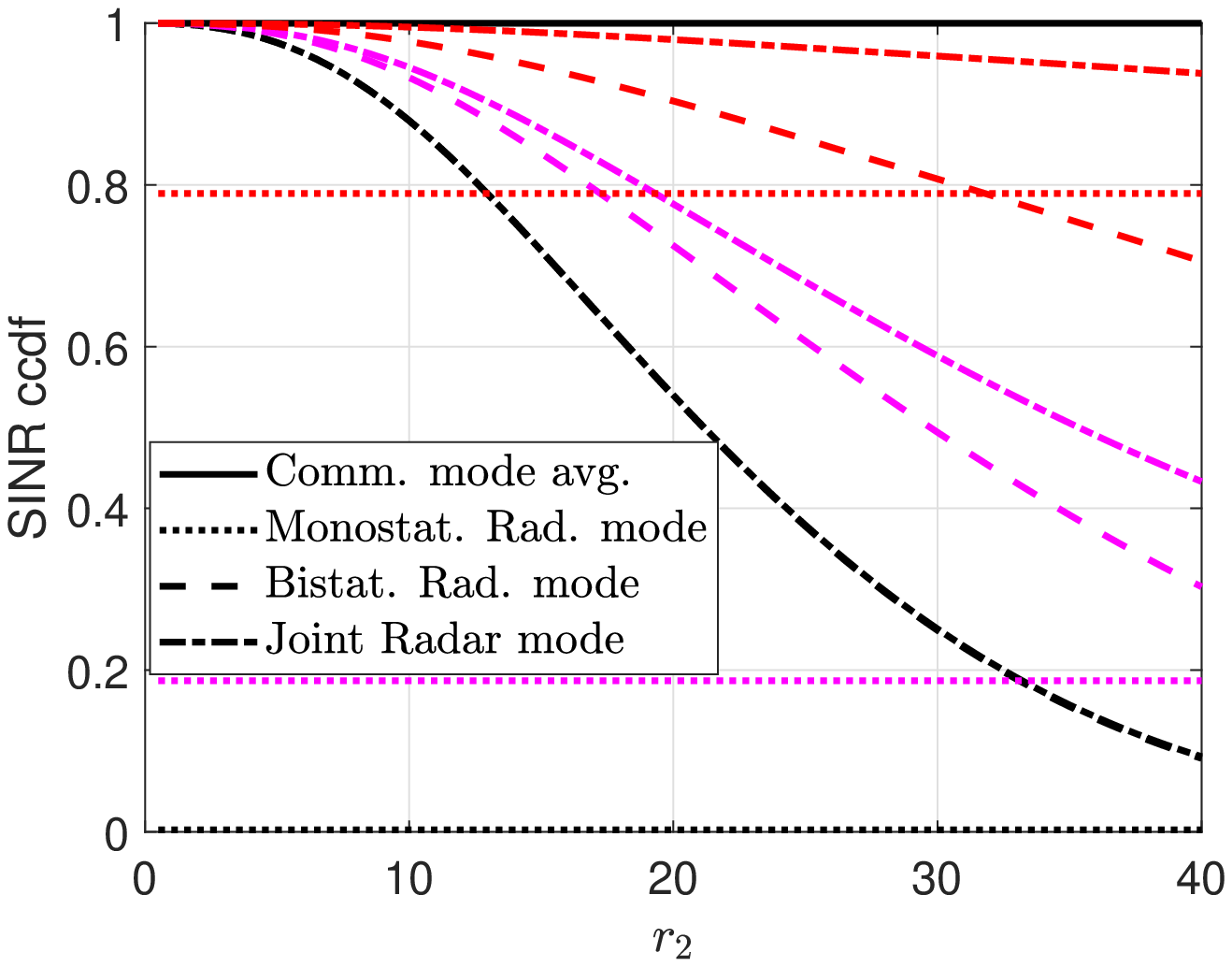}
\subcaption{With DTS}\label{vsR2DTS}
\end{minipage}\;\;\;
\begin{minipage}[htb]{0.47\linewidth}
\centering\includegraphics[width=\linewidth]
{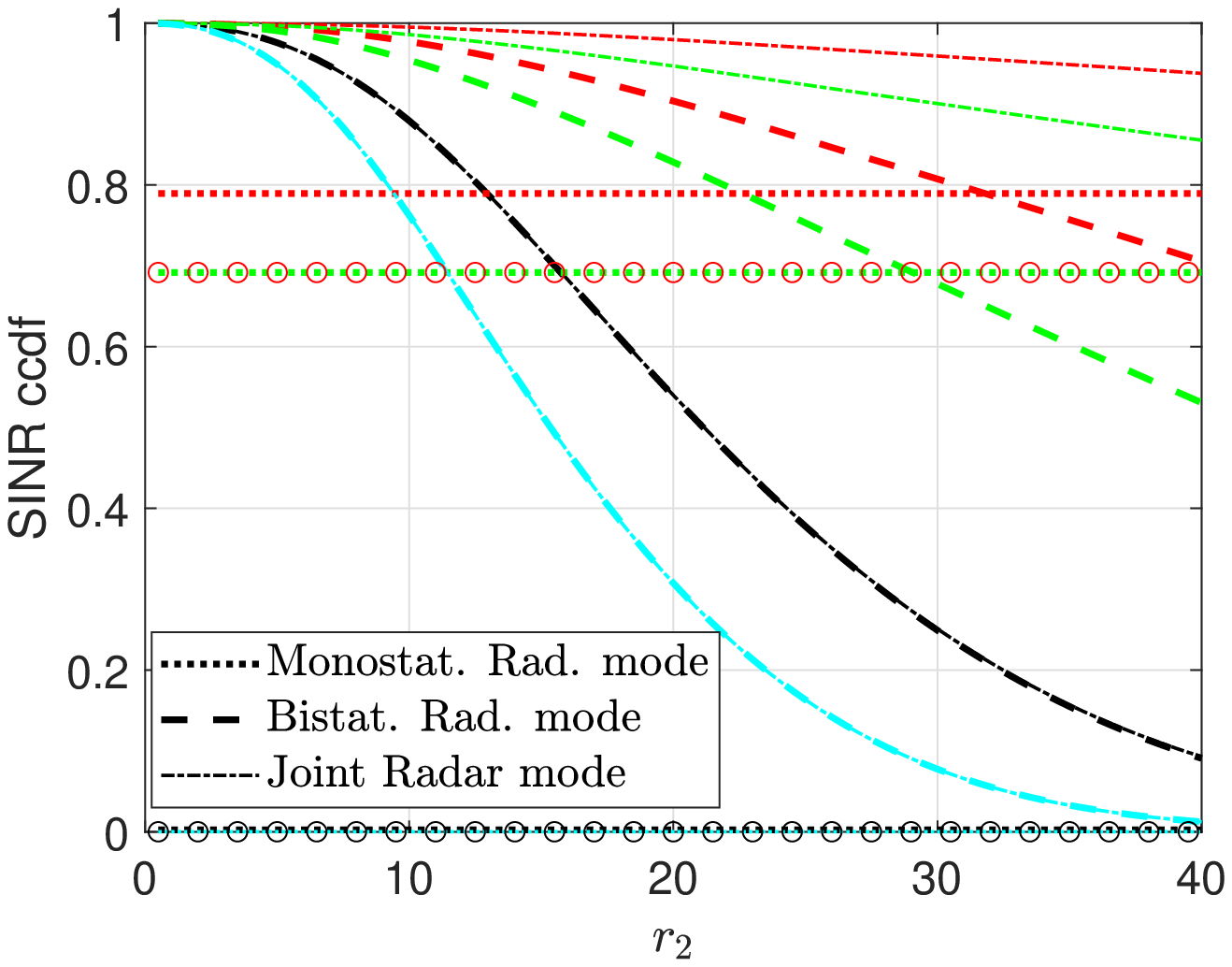}
\subcaption{With and without DTS}\label{vsR2DTSandNoDTS}
\end{minipage}
\caption{Performance of the ISAC network as $r_2$ is increased using $\theta=-40$ dB. Black (magenta, red) lines represent $r_1=15v$ ($r_1=10v$, $r_1=5v$) with DTS using $P_h=5$ and $M=10$. Cyan (green) lines represent $r_1=15v$ ($r_1=5v$) without DTS. Black (red) markers represent the radar-only network with $R_r=15v$ ($R_r=5v$).}\label{vsR2}
\end{figure}

Fig. \ref{vsR2} is a plot of the SINR ccdfs in an ISAC network as $r_2$ is increased. We plot these with and without DTS for the radar-mode for bistatic, monostatic and joint detection as well as the average for the communication-mode. In Fig. \ref{vsR2DTS} we observe that, as anticipated, the SINR ccdf of the communication-mode is not affected by increasing $r_2$. Further, neither is the performance of the monostatic radar-detection as it is not impacted by this link distance. On the other hand, the reliability of detection of the bistatic and joint radar-mode decreases with $r_2$ due to the deteriorating impact of the path loss component $r_2^{-\eta}$. In the case of $r_1=15v$ we observe that since the monostatic detection reliability is very low, the joint detection is equivalent to the bistatic detection in this scenario. Decreasing $r_1$ improves the performance of the bistatic radar-mode as well as the monostatic radar-mode; the joint detection reliability in these scenarios exceeds the bistatic reliability and we see a greater increase in performance of joint detection of the radar-mode than the bistatic case. In Fig. \ref{vsR2DTSandNoDTS}, the radar performance with DTS is always superior to the performance without DTS due to the impact of the stronger signal component dominating the impact of the increased interference in the radar-mode. We also observe that monostatic detection using DTS, without joint detection, has better performance than the radar-only network. Further, joint detection without DTS also has better performance from the radar-only network. While the gain with using DTS is independent of $r_2$, the gain from joint detection without DTS exceeds the gain from using DTS when $r_2$ is smaller. This highlights the improvement of our proposed solutions to enhancing radar performance. It also sheds light on the selection of a solution if only one is available: for smaller $r_2$ a higher gain is obtained by using joint detection and for larger $r_2$ DTS is superior. Of course, using both solutions results in greater gain from the radar-only network than using one solution.

\begin{figure}
\begin{minipage}[htb]{0.95\linewidth}
\centering\includegraphics[width=\linewidth]
{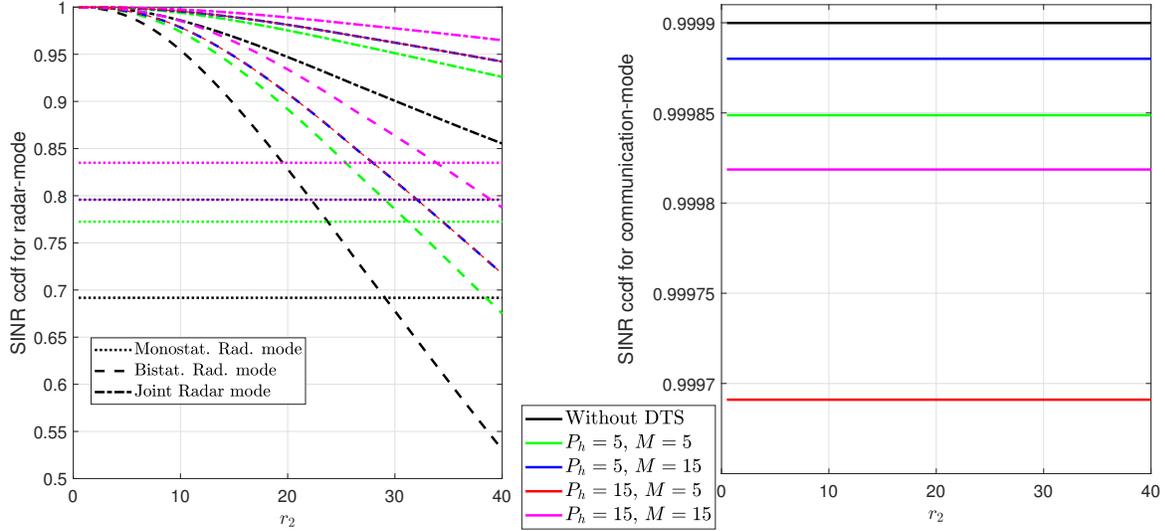}
\subcaption{$P_l=1$}\label{vsR2_radComm_Pl1}
\end{minipage}
\begin{minipage}[htb]{0.95\linewidth}
\centering\includegraphics[width=\linewidth]
{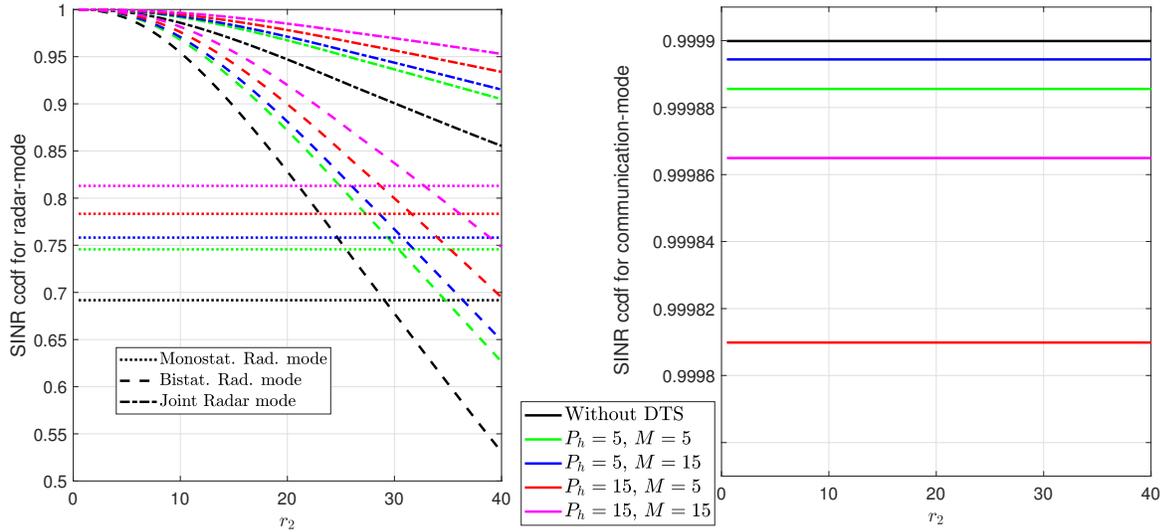}
\subcaption{$P_l=2$}\label{vsR2_radComm_Pl2}
\end{minipage}
\caption{SINR ccdfs of the radar-mode and communication-mode in an ISAC network as $r_2$ is increased using $r_1=15v$ and $\theta=-40$ dB.}\label{vsR2_Pl}
\end{figure}

{Fig. \ref{vsR2_Pl} plots the SINR ccdfs for the radar and communication-modes in an ISAC network vs. $r_2$ again but using different values of $P_h$ and $M$. Increasing each of $P_h$ and $M$ increases the performance of each of the monostatic, bistatic and joint detection in the radar-mode. A very interesting observation in Fig. \ref{vsR2_radComm_Pl1} is that when the values of $P_h$ and $M$ are interchanged, the radar-performance remains unchanged. This was consistently observed to be true and for different values of $\theta$ as well as different values of $P_h$ and $M$ (not plotted for brevity). One implication of this is that, for the radar-mode, the impact of increasing $P_h$ is the same as the impact of increasing $M$, in this scenario, which is not intuitive. On the other hand, for the communication-mode, this is not the case and increasing $M$ improves performance but increasing $P_h$ reduces the performance. This occurs because like for the radar-mode, increasing $M$ reduces interference by decreasing $P_{\rm avg}$. However, increasing $P_h$ only improves the signal in 1 of $M$ slots; the gains from this are less than the increase in $P_{\rm avg}$ and interference caused by increasing $P_h$ in the communication-mode. Thus, unlike the radar-mode, interchanging $P_h$ and $M$ does not result in the same performance for the communication-mode. Instead, for the communication-mode a lower $P_h$ and higher $M$ results in better performance. This highlights an important aspect of parameter selection in ISAC: a certain quality of service for the radar-mode can be obtained using two choices of $M$ and $P_h$ but only one of these choices results in superior performance for the communication-mode.  }

{While increasing $P_h$ and $M$ by the same amount results in the same increase in radar-mode performance when $P_l=1$ in Fig. \ref{vsR2_radComm_Pl1}, we observe in Fig.\ref{vsR2_radComm_Pl2} that this is not the case when $P_l>1$. Increasing $P_h$ has a greater improvement in radar-mode performance than increasing $M$ when $P_l>1$. This stems from the fact that $P_{\rm avg}=\frac{P_h + (M-1)P_l}{M}$, thus, increasing $M$ increases both the numerator and denominator of $P_{\rm avg}$. While the overall impact of increasing $M$ is a reduction in $P_{\rm avg}$, when $P_l>1$, the reduction is lower because in the numerator the increase in $M$ is scaled by $P_l>1$. While $P_h$ impacts both the numerator and denominator of the SINR, it is not scaled by $P_l$ anywhere and the impact of $P_h$ on the signal component of the SINR is more significant; thus, increasing $P_h$ by the same amount improves radar performance more than increasing $M$ when $P_l>1$. In Fig. \ref{vsR2_radComm_Pl2}, we observe that the communication-mode when $P_l>1$ is impacted in the same way as the case for $P_l=1$, i.e., increasing $P_h$ deteriorates performance while increasing $M$ improves performance. For the radar-mode, since increasing $M$ improves performance, although less than increasing $P_h$ by the same amount, it is still worth using a higher $M$ than using higher $P_h$ to improve the performance of both the communication and radar modes.  }




\begin{figure}
\begin{minipage}[htb]{0.49\linewidth}
\centering\includegraphics[width=\linewidth]
{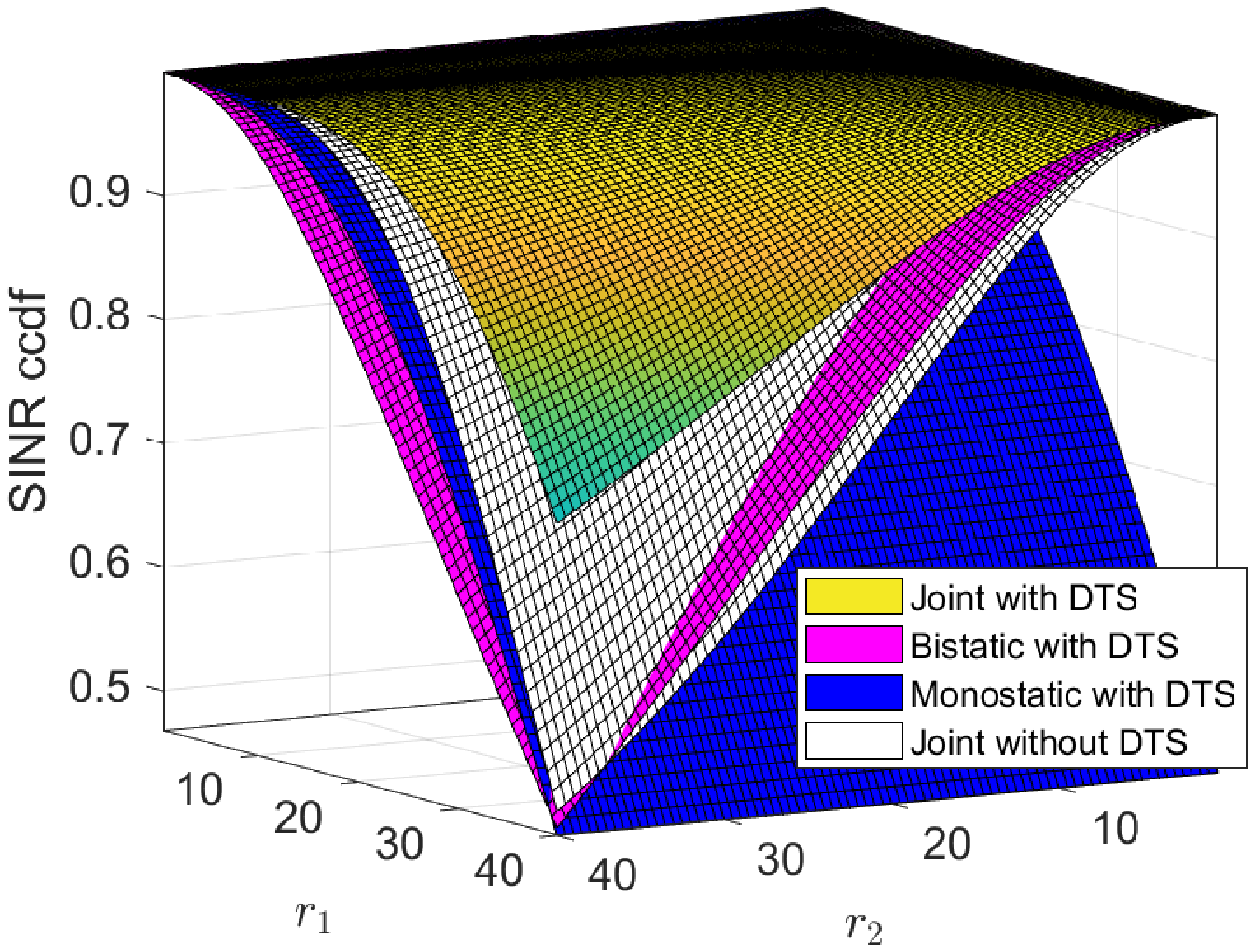}
\caption{Radar-mode performance vs. $r_1$ and $r_2$ using $\theta=-40$ dB. The scenario with DTS uses $P_h=5$ and $M=10$.}\label{radVsR1R2}
\end{minipage}\;\;\;
\begin{minipage}[htb]{0.49\linewidth}
\centering\includegraphics[width=\linewidth]
{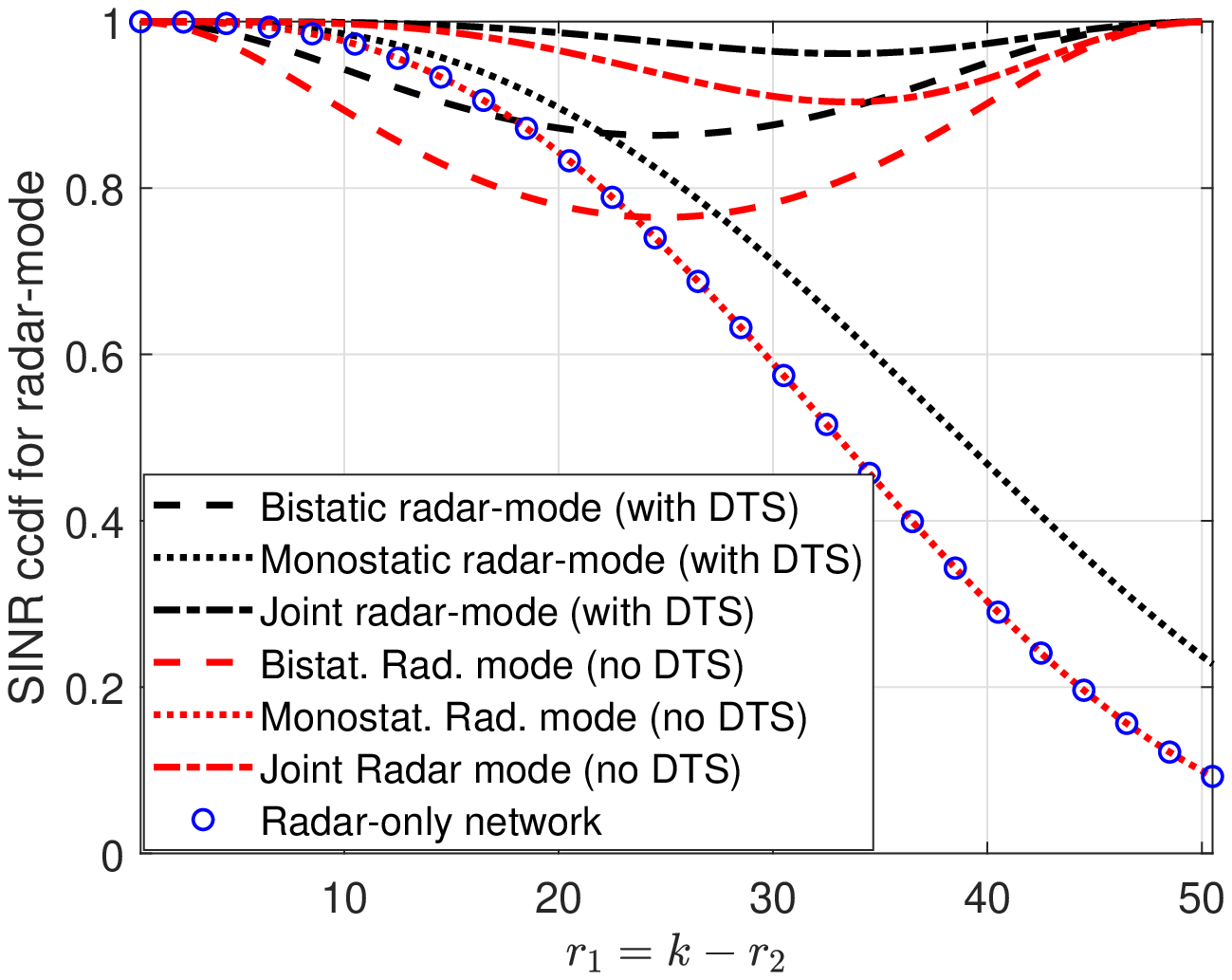}
\caption{Radar-mode performance vs. $r_1=k-r_2$ where $k=50.5$, using $\theta=-40$ dB. The scenario with DTS uses $P_h=5$ and $M=10$.}\label{radVsR1PlusR2}
\end{minipage}
\end{figure}

{In Fig. \ref{radVsR1R2} we plot the SINR ccdf for the bistatic, monostatic and joint radar-mode detection in an ISAC network with DTS vs. $r_1$ and $r_2$. We also plot the joint radar-mode detection for the scenario without DTS. We observe that the joint detection with DTS always outperforms all of the other detection techniques. As anticipated, the surface for monostatic detection is unaffected by $r_2$. We observe that with DTS (and without DTS, not plotted for brevity) the performance of bistatic detection dominates monostatic detection in some regions while the monostatic outperforms bistatic in other regions. Interestingly, we observe that bistatic detection with DTS is able to outperform joint detection without DTS. This observation sheds light on the fact that the DTS has potential to be even more superior to joint detection, highlighting the significance of our proposed strategy for improving the radar-mode performance for ISAC.  }

{In Fig. \ref{radVsR1PlusR2} we plot the SINR ccdf of the radar-mode in an ISAC network vs. $r_1$ where $r_1+r_2=k$. We also plot the performance for the radar-only network. We observe that in the case of monostatic detection in ISAC and in the radar-only network, performance falls monotonously with increasing $r_1$. In the case of bistatic detection, performance first falls and then improves as the decreasing $r_2$ improves the path loss of the tTar-tRad link and therefore performance. This highlights the robustness that bistatic detection in an ISAC network can offer with the deployment of many low-cost passive radars for listening. Further, like bistatic detection, in the case of joint detection, performance first falls and then grows again with $r_1$. {In fact, we observe that joint detection greatly improves robustness to performance deterioration, reducing the variation in performance with link distances significantly. As anticipated, employing the DTS improves performance and in the case of joint detection, further reduces variation in performance with link distance resulting in near-constant performance for the plotted scenario.} {For the plotted scenario, at the performance minima for each, with (without) DTS, bistatic detection is 279.3 \% (726.2 \%) higher than monostatic detection, joint detection is 11.4 \% (18.1 \%) higher than bistatic detection and joint detection is 322.4 \% (876.2 \%) higher than monostatic detection. Similarly, at the performance minima for each, with (without) DTS, performance using bistatic detection is up to 832.9 \% (726.2 \%) higher than a radar-only network, using joint detection is up to 939 \% (876.2 \%) higher than a radar-only network, and even with just DTS monostatic detection is up to 146 \% higher than a radar-only network.} These results highlight that employing joint detection and the DTS are excellent and practical solutions for improving radar-mode performance in an ISAC network and that with ISAC, significant performance improvement from a radar-only network can be obtained. }


\begin{figure}
\begin{minipage}[htb]{0.47\linewidth}
\centering\includegraphics[width=\linewidth]
{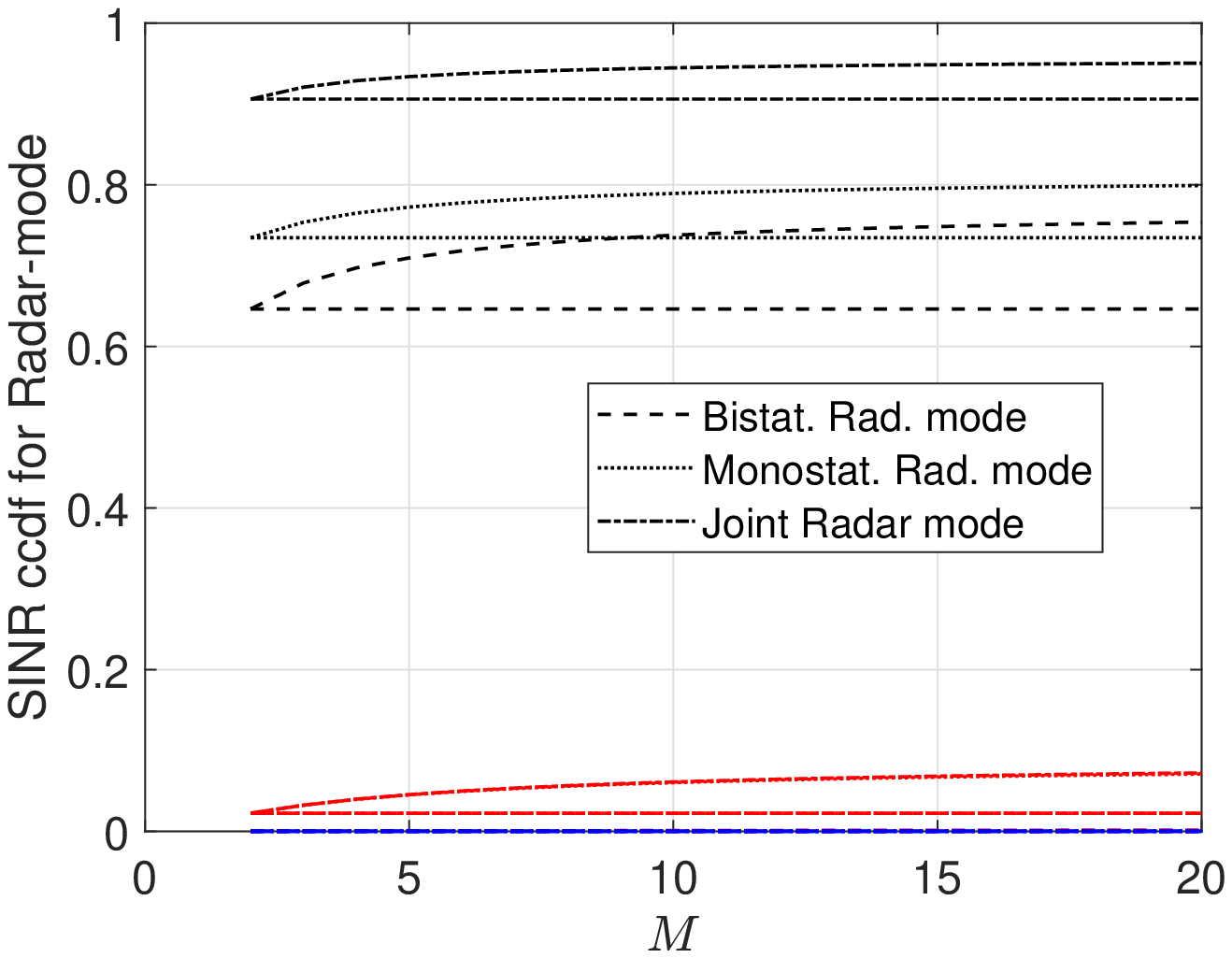}
\subcaption{Radar-mode. Curves (horizontal lines) are for the network employing DTS (without DTS). Black (red, blue) represent $\theta=-40$ dB ($\theta=-10$ dB, $\theta=10$ dB) }\label{radVsM}
\end{minipage}\;\;\;
\begin{minipage}[htb]{0.47\linewidth}
\centering\includegraphics[width=\linewidth]
{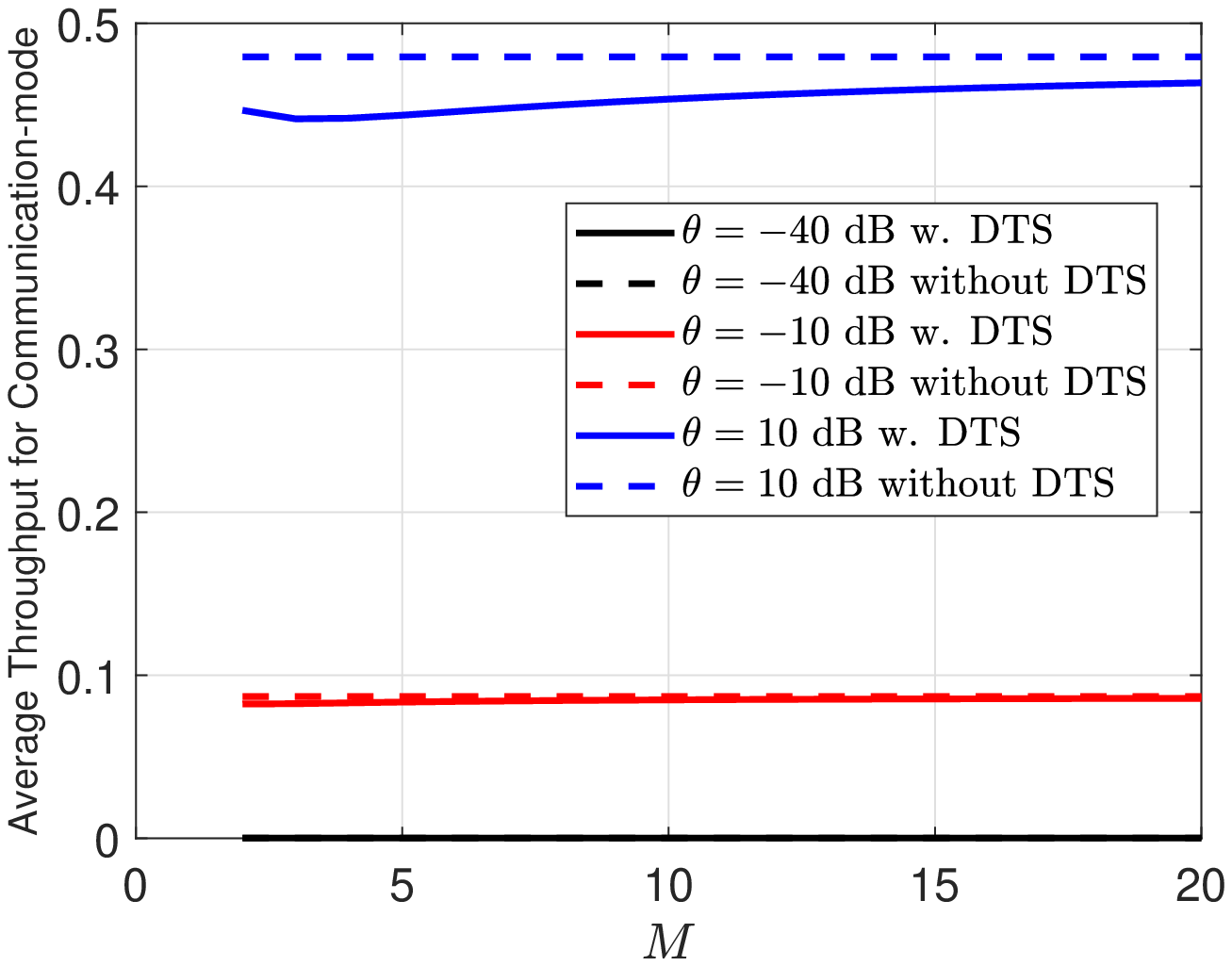}
\subcaption{Communication-mode.\\$\;$\\$\;$}\label{commVsM}
\end{minipage}
\caption{Performance vs. $M$ using $r_1=5v$ and $r_2=7v$. The scenario with DTS uses $P_h=5$.}\label{vsM}
\end{figure}

In Fig. \ref{vsM} we plot the SINR ccdfs of the radar-mode and the average throughput of the communication-mode as performance metrics. Note that for the plotted values, the SINR ccdf decreases with $\theta$ while the throughput increases. For the radar-mode in Fig. \ref{radVsM}, as anticipated, the performance with DTS is superior to that without DTS for each of the bistatic, monostatic and joint detection cases. Further, increasing $M$ improves radar performance with DTS for each detection technique as $P_{\rm avg}$ and therefore interference decrease. For each detection technique, the performance starts saturating at large $M$ as $P_{\rm avg} \to P_l$, which is a constant, as $M$ grows. As anticipated, the performance without DTS is unaffected by $M$. Note that the performance of the radar-only network is equivalent to monostatic detection without DTS. For the values of $r_2$ in Fig. \ref{vsM}, bistatic detection is worse than monostatic and therefore the radar-only network; however, even in this scenario we observe that with joint detection, both with and without DTS, a significant improvement in performance is observed from the radar-only network.


In Fig. \ref{commVsM} the communication-mode throughput is plotted and we observe that the performance in an ISAC network without DTS, which is equivalent to a traditional cellular network without ISAC, is superior to that with DTS due to the increase in interference that DTS causes. {Unlike the radar-mode, the performance difference for the communication-mode between the scenario with and without DTS does not monotonously increase with $M$.} In fact, we observe the existence of $M$ where there is a throughput minima with DTS. This occurs because the communication-mode with DTS faces a trade off between the benefit from the transmission in the time slot with higher power $P_h$ and higher interference caused by DTS. When $M=2$, the gains from the slot with $P_h$ dominates the deterioration from the increased interference. As $M$ grows, the impact of the gains from the high power transmission decreases but so does $P_{\rm avg}$ and interference. At first, the impact of the decrease in gains is more and we see a drop in performance with $M$. After this, the impact of the decreasing $P_{\rm avg}$ and interference dominates and the communication-mode performance increases with $M$. This highlights the significance of careful selection of parameters such as $M$ to optimize performance. {Note that as $M$ grows the performance of the communication-mode slowly grows. This indicates that with high $M$ we can obtain good radar-mode as well as communication-mode performance. However, it ought to be mentioned that while high $M$ improves the \emph{quality} of radar detection, larger $M$ reduces the \emph{quantity} of radar detection. Parameter selection thus needs to be done as a function of both radar and communication mode quality requirements but also needs to take into account the sensing requirements of the network.}

\begin{figure}
\begin{minipage}[htb]{0.315\linewidth}
\centering\includegraphics[width=\linewidth]
{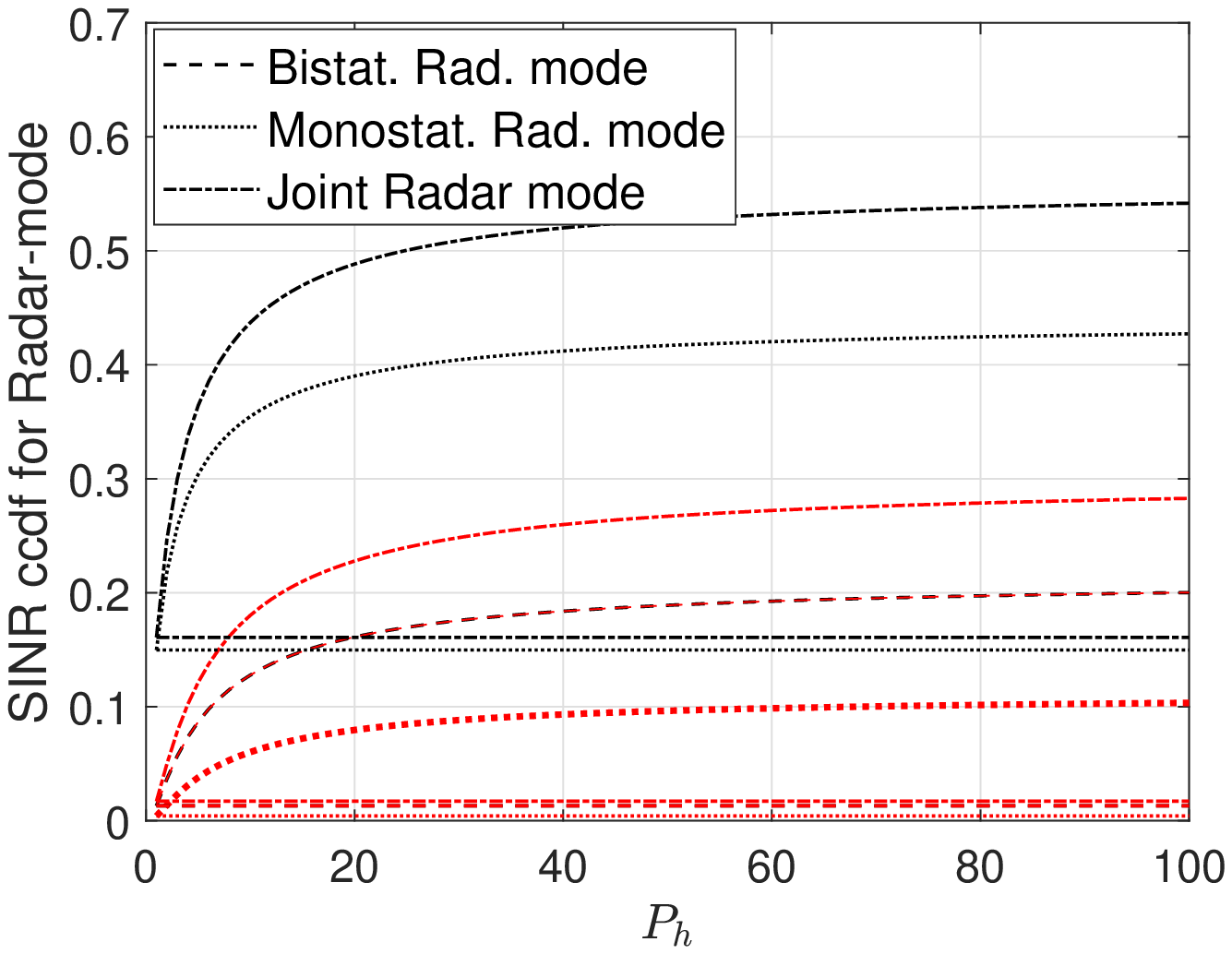}
\subcaption{Radar-mode with $\theta=-20$ dB. Black (red) lines represent $r_1=5v$ and $r_2=7v$ ($r_1=7v$ and $r_2=5v$).\\  }\label{radVsPh_diffRs}
\end{minipage}\;\;
\begin{minipage}[htb]{0.315\linewidth}
\centering\includegraphics[width=\linewidth]
{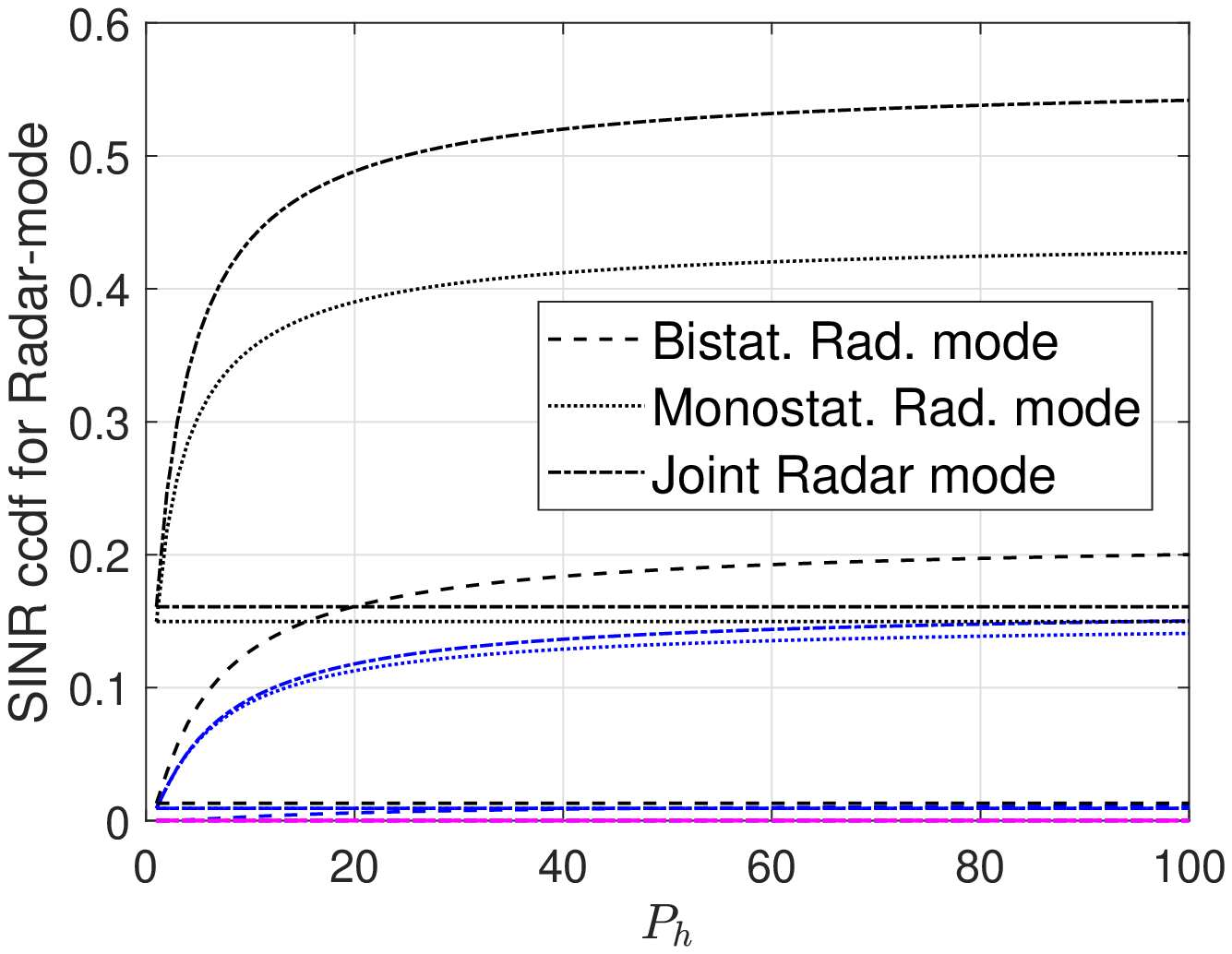}
\subcaption{Radar-mode with DTS using $r_1=5v$ and $r_2=7v$. Black (blue, magenta) lines represent $\theta=-20$ dB ($\theta=-10$ dB, $\theta=10$ dB).}\label{radVsPh_diffTheta}
\end{minipage}\;\;
\begin{minipage}[htb]{0.315\linewidth}
\centering\includegraphics[width=\linewidth]
{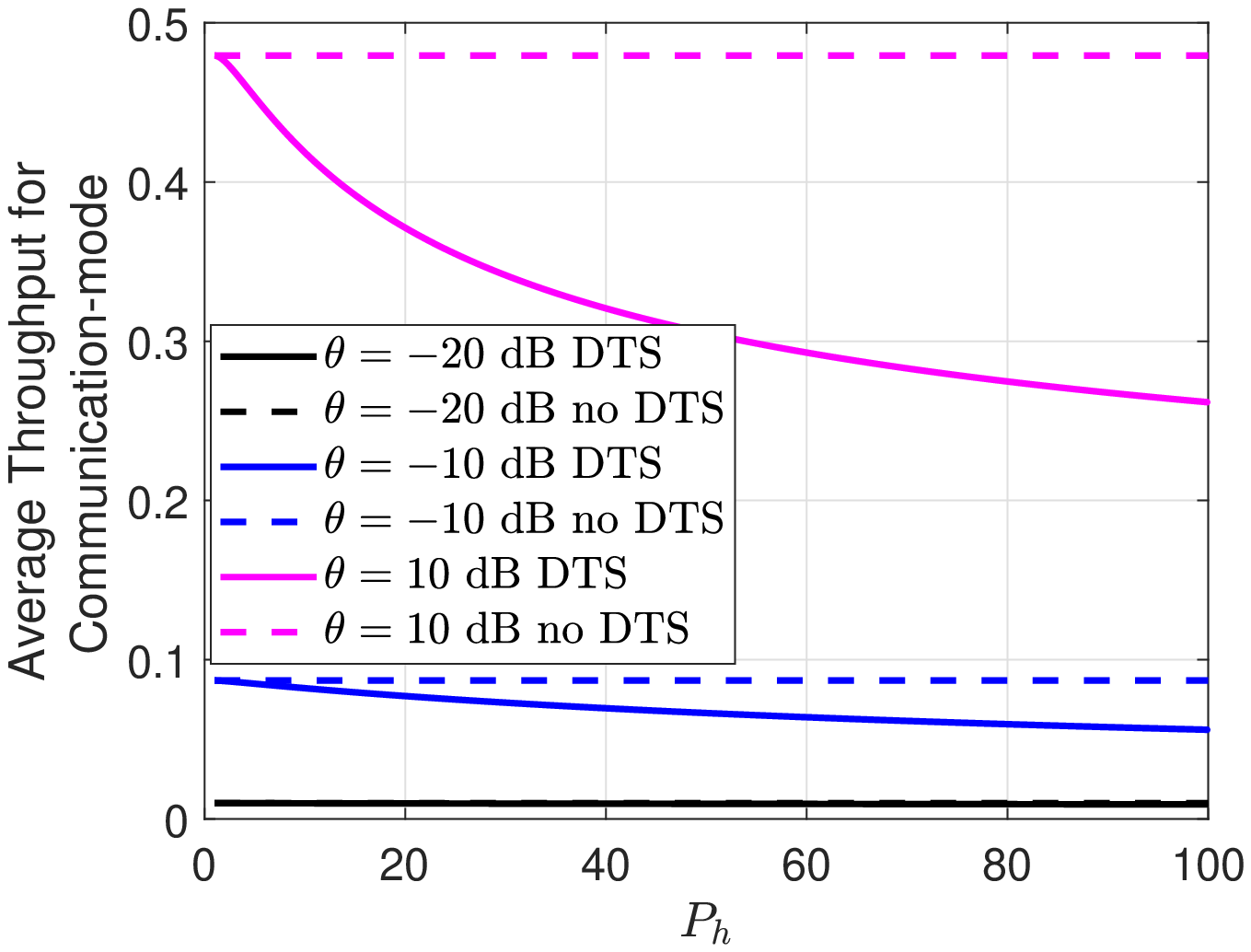}
\subcaption{Communication-mode.\\ $\;$\\$\;$\\$\;$\\}\label{commVsPh}
\end{minipage}
\caption{Performance of the ISAC network vs. $P_h$. Curves (horizontal lines) are for DTS (without DTS). For the network with DTS, $M=10$.}\label{vsPh}
\end{figure}

Fig. \ref{vsPh} plots the performance of both the communication and radar modes as $P_h$ is increased. Increasing $P_h$ has a trade off for both modes as the signal in slot 1 of the communication-mode and the radar-mode's signal power increase, but so does the average interference. In Figs. \ref{radVsPh_diffRs} and \ref{radVsPh_diffTheta} we observe that with increasing $P_h$ performance of the radar-mode with each of bistatic, monostatic and joint detection grows at first as the benefit from the increasing signal power dominates in this regime. However, at higher $P_h$ the performance of the radar-mode starts to saturate for each detection technique. This occurs due to $P_{\rm avg} \to P_h/M$ at large $P_h$. As the network being considered is large, the impact of the noise is not significant compared to the interference and an interference-limited regime can be assumed (i.e., $\text{SINR} \to \text{SIR}$). At high $P_h$, the $P_h$ term in the signal component and interference component cancel out and the performance of the radar-mode starts to saturate to a constant with increasing $P_h$. Note that this performance is still superior to a radar-only network as the SINR in ISAC with DTS is amplified by $M$. As anticipated, the performance of the ISAC network without DTS is a lower bound on the performance with DTS and is unaffected by $P_h$. We also observe in Fig. \ref{radVsPh_diffRs} that interchanging the values of $r_1$ and $r_2$ (or swapping the positions of the tBS and tTar) does not impact the performance of bistatic detection as the double path loss in this scenario remains the same. However, interchanging $r_1$ and $r_2$ has a significant impact on monostatic detection and consequently joint detection. In particular, we observe that when $r_1$ is assigned the smaller value, monostatic and joint detection improve significantly. {This sheds light on the significance of the distance $r_1$ which plays a larger role on radar-mode performance in ISAC than $r_2$. It highlights important aspects of decision making such as using monostatic radar detection only when searching for targets within a certain vicinity. Another solution this highlights is to use ISAC for targets near the cell center and using traditional radar detection for cell edge targets.}


In Figs. \ref{radVsPh_diffTheta} and \ref{commVsPh} the radar-mode reliability of detection and the average communication-mode throughput are plotted vs. $P_h$ for different $\theta$. As expected, increasing $\theta$ reduces the radar-mode reliability for the bistatic, monostatic and joint detection in Fig. \ref{radVsPh_diffTheta}. The communication-mode throughput on the other hand increases with $\theta$ in Fig. \ref{commVsPh}. The performance of the communication-mode without DTS upper bounds the average performance with DTS. We observe that with DTS, the performance at first decreases slowly with $P_h$ as the impact of the improving signal, while still not dominant, has a larger impact. As $P_h$ grows, the impact of increasing $P_h$ on increasing interference becomes increasingly significant, as the signal with $P_h$ for the communication-mode is only in 1 of $M$ slots, and the performance decreases more rapidly with $P_h$. At higher $P_h$ when $P_{\rm avg} \to P_h/M$, the throughput saturates. 


\section{Conclusion}

In this work we have studied a large network employing ISAC where a single transmit signal by the BS serves both the radar and communication modes. Target detection in the radar-mode of ISAC is done via monostatic detection at the BS or bistatic detection using passive radars. While the communication-mode UEs have a direct link between the transmitting BS and receiving UE, the radar-mode performance is significantly more vulnerable due to the double path-loss in the signal component; this, in addition to direct links from interferers worsen the radar-mode SINR further. To combat this, we propose two solutions. First, we propose using joint monostatic and bistatic detection via cooperation at the BS. We study and analyze the performance of bistatic, monostatic and joint detection in a large ISAC network. Second, we propose a novel dynamic transmission strategy (DTS) to improve radar performance. Since detection may not be required in every time slot, each BS dynamically transmits with higher power $P_h$ in one of $M$ time slots, while it transmits with lower power $P_l$ in the remaining slots. Since the slot with $P_h$ is chosen at random for each BS, the average power of an interferer is $P_{\rm avg}$. This way radar detection is done in the slot with $P_h$ and since $P_{\rm avg}<P_h$, the reduced interference further aids the radar-mode with DTS. While the DTS reduces the quantity of radar detection by $1/M$, we find that it improves quality significantly. The communication-mode faces a trade off with DTS between higher transmit power in one slot improving its performance and the increase in average interference deteriorating performance in the remaining slots. Our results show that overall the DTS hurts the communication-mode due to higher network interference. However, the deterioration of the communication-mode is significantly lower than the improvement of the radar-mode. {We show that with careful choice of $M$ we can reduce the deterioration of the communication-mode and improve the performance of the radar-mode at the expense of reduced quantity of detection.} {We find that in certain scenarios the same radar-mode performance can be obtained with two choices of DTS parameters, while only one of these results in superior performance for the communication-mode, highlighting the significance of parameter selection in an ISAC network with DTS.} We also find that monostatic detection is the superior solution in certain scenarios like cell-center targets, while the bistatic detection is superior in other scenarios such as farther off targets. {Our results show that bistatic detection with dense deployment of low-cost passive radars for listening can improve the robustness of an ISAC network's target detection capability. Further,} joint detection, as anticipated, improves performance significantly more than either one detection technique in most situations; however, in certain scenarios where one of monostatic or bistatic dominates the other, joint detection may not be required. {Joint detection with DTS also reduces the variation in quality of target detection as the distance between the BS and target increases.} Overall, we find that joint detection and DTS are excellent and practical solutions to improve radar-mode performance in ISAC networks under different scenarios and can significantly improve radar-mode performance from that of a traditional radar-network.





%


\bibliographystyle{IEEEtran}
\bibliography{References}

\begin{thebibliography}{10}
\providecommand{\url}[1]{#1}
\csname url@samestyle\endcsname
\providecommand{\newblock}{\relax}
\providecommand{\bibinfo}[2]{#2}
\providecommand{\BIBentrySTDinterwordspacing}{\spaceskip=0pt\relax}
\providecommand{\BIBentryALTinterwordstretchfactor}{4}
\providecommand{\BIBentryALTinterwordspacing}{\spaceskip=\fontdimen2\font plus
\BIBentryALTinterwordstretchfactor\fontdimen3\font minus
  \fontdimen4\font\relax}
\providecommand{\BIBforeignlanguage}[2]{{%
\expandafter\ifx\csname l@#1\endcsname\relax
\typeout{** WARNING: IEEEtran.bst: No hyphenation pattern has been}%
\typeout{** loaded for the language `#1'. Using the pattern for}%
\typeout{** the default language instead.}%
\else
\language=\csname l@#1\endcsname
\fi
#2}}
\providecommand{\BIBdecl}{\relax}
\BIBdecl

\bibitem{jrc_overview0}
J.~A. Zhang, M.~L. Rahman, K.~Wu, X.~Huang, Y.~J. Guo, S.~Chen, and J.~Yuan,
  ``Enabling joint communication and radar sensing in mobile networks—a
  survey,'' \emph{IEEE Commun. Surveys and Tutorials}, vol.~24, no.~1, pp.
  306--345, 2022.

\bibitem{jrc_overview1}
J.~A. Zhang, F.~Liu, C.~Masouros, R.~W. Heath, Z.~Feng, L.~Zheng, and
  A.~Petropulu, ``An overview of signal processing techniques for joint
  communication and radar sensing,'' \emph{IEEE J. Selec. Topics in Sig.
  Proc.}, vol.~15, no.~6, pp. 1295--1315, 2021.

\bibitem{jrc_sg6}
F.~D.~S. Moulin, C.~Oestges, and L.~Vandendorpe, ``Characterisation and
  cancellation of interference with multiple phase-coded {FMCW} dual-function
  {RADAR} communication systems,'' in \emph{Proc. of IEEE 95th Vehicular
  Technology Conference (VTC22)}, 2022, pp. 1--7.

\bibitem{12challenges}
M.~Chafii, L.~Bariah, S.~Muhaidat, and M.~Debbah, ``Twelve scientific
  challenges for {6G}: Rethinking the foundations of communications theory,''
  \emph{IEEE Commun. Surveys and Tutorials}, pp. 1--1, 2023.

\bibitem{jrc_sg4}
S.~Ram, S.~Singhal, and G.~Ghatak, ``Optimization of network throughput of
  joint radar communication system using stochastic geometry,'' \emph{Frontiers
  in Signal Processing}, vol.~2, 04 2022.

\bibitem{jrc_irs1}
S.~Yan, S.~Cai, W.~Xia, J.~Zhang, and S.~Xia, ``A reconfigurable intelligent
  surface aided dual-function radar and communication system,'' in \emph{Proc.
  of IEEE International Symposium on Joint Communications \& Sensing
  (JC\&S22)}, 2022, pp. 1--6.

\bibitem{Bazzi_JRC1}
A.~Bazzi and M.~Chafii, ``On outage-based beamforming design for
  dual-functional radar-communication {6G} systems,'' \emph{IEEE Trans.
  Wireless Commun.}, pp. 1--1, 2023.

\bibitem{Bazzi_JRC2}
------, ``On integrated sensing and communication waveforms with tunable
  {PAPR},'' \emph{IEEE Trans. Wireless Commun.}, pp. 1--1, 2023.

\bibitem{my_nomaMag}
K.~S. Ali, H.~Elsawy, A.~Chaaban, and M.~S. Alouini, ``Non-orthogonal multiple
  access for large-scale {5G} networks: Interference aware design,'' \emph{IEEE
  Access}, vol.~5, pp. 21\,204--21\,216, 2017.

\bibitem{MH_Book2}
B.~Blaszczyszyn, M.~Haenggi, P.~Keeler, and S.~Mukherjee, \emph{Stochastic
  Geometry Analysis of Cellular Networks}.\hskip 1em plus 0.5em minus
  0.4em\relax Cambridge University Press, 2018.

\bibitem{h_tut}
H.~ElSawy, A.~Sultan-Salem, M.~S. Alouini, and M.~Z. Win, ``Modeling and
  analysis of cellular networks using stochastic geometry: A tutorial,''
  \emph{IEEE Commun. Surveys and Tutorials}, vol.~19, no.~1, pp. 167--203,
  Firstquarter 2017.

\bibitem{myNOMA_tcom}
K.~S. Ali, M.~Haenggi, H.~E. Sawy, A.~Chaaban, and M.~Alouini, ``Downlink
  non-orthogonal multiple access {(NOMA)} in {P}oisson networks,'' \emph{IEEE
  Trans. Commun.}, vol.~67, no.~2, pp. 1613--1628, Feb. 2019.

\bibitem{mySecrecy}
K.~S. {Ali}, H.~{ElSawy}, M.~{Haenggi}, and M.~{Alouini}, ``The effect of
  spatial interference correlation and jamming on secrecy in cellular
  networks,'' \emph{IEEE Wireless Comm. Letters}, vol.~6, no.~4, pp. 530--533,
  2017.

\bibitem{myPartialNOMA}
K.~S. {Ali}, E.~{Hossain}, and M.~J. {Hossain}, ``Partial non-orthogonal
  multiple access ({NOMA}) in downlink {P}oisson networks,'' \emph{IEEE Trans.
  Wireless Commun.}, vol.~19, no.~11, pp. 7637--7652, 2020.

\bibitem{jrc_sg1}
P.~Ren, A.~Munari, and M.~Petrova, ``Performance tradeoffs of joint
  radar-communication networks,'' \emph{IEEE Wireless Comm. Letters}, vol.~8,
  no.~1, pp. 165--168, 2019.

\bibitem{jrc_sg2}
Z.~Fang, Z.~Wei, Z.~Feng, X.~Chen, and Z.~Guo, ``Performance of joint radar and
  communication enabled cooperative detection,'' in \emph{Proc. of IEEE
  International Conference on Communications in China (ICCC19)}, 2019, pp.
  753--758.

\bibitem{jrc_sg3}
D.~Ghozlani, A.~Omri, S.~Bouallegue, H.~Chamkhia, and R.~Bouallegue,
  ``Stochastic geometry-based analysis of joint radar and communication-enabled
  cooperative detection systems,'' in \emph{Proc. of International Conference
  on Wireless and Mobile Computing, Networking and Communications (WiMob21)},
  2021, pp. 325--330.

\bibitem{jrc_sg5}
\BIBentryALTinterwordspacing
A.~Salem, C.~Masouros, F.~Liu, and D.~López-Pérez, ``Rethinking dense cells
  for integrated sensing and communications: A stochastic geometric view,''
  \emph{CoRR}, vol. abs/2212.12942, 2022. [Online]. Available:
  \url{https://arxiv.org/abs/2212.12942}
\BIBentrySTDinterwordspacing

\bibitem{2sic2_fd5}
M.~Duarte and A.~Sabharwal, ``Full-duplex wireless communications using
  off-the-shelf radios: Feasibility and first results,'' in \emph{Proc. of the
  Forty Fourth Asilomar Conference on Signals, Systems and Computers
  (ASILOMAR10)}, Nov. 2010, pp. 1558--1562.

\bibitem{1sic1_fd5}
\BIBentryALTinterwordspacing
J.~I. Choi, M.~Jain, K.~Srinivasan, P.~Levis, and S.~Katti, ``Achieving single
  channel, full duplex wireless communication,'' in \emph{Proc. of the
  Sixteenth Annual International Conference on Mobile Computing and Networking
  (MobiCom10)}, 2010, pp. 1--12. [Online]. Available:
  \url{http://doi.acm.org/10.1145/1859995.1859997}
\BIBentrySTDinterwordspacing

\bibitem{myD2DFD}
K.~S. Ali, H.~ElSawy, and M.~S. Alouini, ``Modeling cellular networks with
  full-duplex {D2D} communication: A stochastic geometry approach,'' \emph{IEEE
  Trans. Commun.}, vol.~64, no.~10, pp. 4409--4424, Oct. 2016.

\bibitem{FD_JRCmag}
C.~B. Barneto, S.~D. Liyanaarachchi, M.~Heino, T.~Riihonen, and M.~Valkama,
  ``Full duplex radio/radar technology: The enabler for advanced joint
  communication and sensing,'' \emph{IEEE Wireless Communications}, vol.~28,
  no.~1, pp. 82--88, 2021.

\bibitem{mh_ue_PP}
M.~Haenggi, ``User point processes in cellular networks,'' \emph{IEEE Wireless
  Comm. Letters}, vol.~6, no.~2, pp. 258--261, Apr. 2017.

\bibitem{sg_radar3}
A.~Al-Hourani, R.~J. Evans, S.~Kandeepan, B.~Moran, and H.~Eltom, ``Stochastic
  geometry methods for modeling automotive radar interference,'' \emph{IEEE
  Trans. Intelligent Transportation Sys.}, vol.~19, no.~2, pp. 333--344, 2018.

\bibitem{radar_scnrRef1}
S.~S. Ram, G.~Singh, and G.~Ghatak, ``Optimization of radar parameters for
  maximum detection probability under generalized discrete clutter conditions
  using stochastic geometry,'' \emph{IEEE Open Jrnl. Sig. Proc.}, vol.~2, pp.
  571--585, 2021.

\bibitem{Alzer}
\BIBentryALTinterwordspacing
H.~Alzer, ``On some inequalities for the incomplete gamma function,''
  \emph{Mathematics of Computation}, vol.~66, no. 218, pp. 771--778, 1997.
  [Online]. Available: \url{http://www.jstor.org/stable/2153894}
\BIBentrySTDinterwordspacing

\bibitem{mhaenggi_Book}
M.~Haenggi, ``Stochastic geometry for wireless networks,'' \emph{Cambridge
  University Press}, 2012.

\end{thebibliography}

\end{document}